# Low-threshold RNGH instabilities in quantum cascade lasers

Nikola Vukovic[1], Jelena Radovanovic[1], Vitomir Milanovic[1] and Dmitri L. Boiko[2]

[1] *University of Belgrade, School of Electrical Engineering,*
*Bulevar kralja Aleksandra 73, 11000 Belgrade, Serbia*
*e-mail address: nikolavukovic89@gmail.com*
[2] *Centre Suisse d'Électronique et de Microtechnique (CSEM),*
*Jaquet-Droz 1, CH-2002 Neuchâtel, Switzerland*
*e-mail address: dmitri.boiko@csem.ch*

A theoretical study on low-threshold multimode instabilities in quantum cascade lasers (QCLs) and laser diodes (LDs) is presented. Previously, low threshold Risken-Nummedal-Graham-Haken (RNGH) instabilities were reported in several experimental studies of QCLs. They were attributed so far to the combined effect of the induced grating of carrier population and of a built-in saturable absorption feature that may be present in the monolithic single-section cavity of these Fabry-Pérot lasers. Here we show that low-threshold RNGH instabilities in QCLs occur due to a combined effect of the carrier coherence grating and carrier population grating induced in the gain medium and not due to an intracavity saturable absorption. We find that QCLs with a few mm long cavity exhibit intermittent RNGH self-pulsations while regular self-pulsations are possible in short-cavity QCLs, with the cavity length of 100 μm or smaller. We examine a transient behavior to RNGH self-pulsations in short-cavity QCLs and find features that resemble cooperative superradiance. Our findings open a practical way of achieving ultra-short pulse production regimes in the mid-infrared spectral range. Applying same approach to semiconductor laser diodes (LDs) we explain the absence of RNGH self-pulsation in single-section LDs based on a quantum well gain media, while practically established method for reaching the ultrafast coherent emission regimes in LDs is to incorporate a separately contacted saturable absorber section in the LD cavity.

# I. INTRODUCTION

Following the first demonstration at Bell Labs in 1994 by F. Capasso and J. Faist, quantum cascade lasers (QCLs) have experienced rapid and dramatic improvements in power, efficiency and wavelength span. Continuously operating QCLs in the mid infra-red (MIR) spectral range have important applications in security, health, gas sensing and gas analysis [1]. Production of frequency combs in QCLs has been demonstrated and proven to be very promising for spectroscopic applications [2]. Yet, QCL potential as a versatile spectroscopic tool could be significantly enhanced if operation in the ultra-short pulse regime will be possible. Such regime will enable time resolved spectroscopic measurements in MIR and FIR ranges for a wide spread of applications such as LIDARs (LIght Detection and Ranging), Earth observation, environmental remote sensing of molecules, in particular, of greenhouse gases, high-speed QCL based communications utilizing the atmospheric transmission windows of 3-5 μm or 8-12 μm and many more [1].

However ultrafast carrier relaxation at picosecond time scale prohibits passive mode-locking or Q-switching operation in QCLs because of the cavity roundtrip time being longer than the gain recovery time [3]. Therefore gain switched pulse production has been attempted, yielding 120 ps pulse width [4]. On the other hand, active mode-locking was achieved in QCLs utilizing diagonal transition, with the upper state lifetime being increased to 50 ps so as to match the cavity roundtrip time. Yet the diagonal transition renders QCL operation temperature out of the practical use [5].

One more promising approach to generate short MIR pulses has emerged from experimental observations that some of CW operating QCL samples exhibit features of low-threshold [6] Risken-Nummedal-Graham-Haken (RNGH) multimode instabilities [7,8] and hence they might be capable of producing ultra-fast self-pulsations. However, the second-order interferometric autocorrelation measurements have shown that the output optical pulses have significant stochastic constituent [5], if these autocorrelation traces should not be attributed at all to coherence spikes. In order to confirm RNGH instability and to understand conditions for reaching regular self-pulsation in QCLs, one should figure out the origin of such low second threshold in these lasers.

In general, the multimode RNGH instabilities are related to a non-adiabatic [9] behavior of the medium polarization, excitation of rapid coupled oscillations of the medium polarization $P$ and population inversion $N$. They show up as self-pulsations in the output laser emission, whereas the optical spectrum is expected to be split in two mode clusters (sidebands) with frequency separation of the order of the Rabi oscillation frequency.

Below the RNGH instability threshold, the optical mode is fully controlled by the laser cavity. The buildup of Rabi oscillations at above the second threshold indicates that now a medium macroscopic polarization (coherence) has the major impact on the optical field in the laser cavity. The original RNGH theory, which was established for a CW operating *unidirectional ring laser*, requires the pump rates of at least 9 times above the lasing threshold in order for multimode instability to occur.

However the experiments revealed a different second threshold behavior. In Ref [10], a ring-cavity dye laser with homogeneously broadened gain line has shown RNGH-like multimode instability at small excess above the lasing threshold. More specifically, quenching of the main lasing mode and excitation of two optical sidebands was observed at 1.2 times above the lasing threshold. The frequency separation of the sidebands was close to the Rabi oscillation frequency or one of its sub-harmonics and it was increasing with the pump rate. Later on there were several claims on RNGH instabilities in CW operating $Er^{3+}$ doped ring fiber lasers [11-13], and ring Nd:YAG lasers [14]. However, interpretation of these experimental observations remained doubtful. Thus the period of self-pulsations was close to the cavity roundtrip time, so they can be attributed to mode-locking phenomena. Most unexpectedly, the self-pulsations were occurring almost at the lasing threshold, contrary to the theoretical predictions for at least 9-fold excess above the lasing threshold. Note that the relaxation $T_1$ and dephasing $T_2$ times in the active gain media of these lasers are relatively long as compared to their semiconductor counterparts and hence their Rabi oscillation frequencies are low. It was understood that (i) the cavity length must be made very long in order to match the intermodal frequency separation to that one of the Rabi oscillations, and that (ii) the lowering of second threshold in $Er^{3+}$ doped ring fiber lasers can be attributed to a 3-level structure of the optical transitions in $Er^{3+}$ ions [15]. Nevertheless the fact of experimental observation of RNGH instability remained debatable.

Ten years later, a clear Rabi frequency splitting between two clusters of modes was observed in the lasing spectra of CW operating QCLs [6], indicating multimode RNGH instability in these *Fabry-Pérot (FP) cavity lasers*. Interestingly, the RNGH instability was observed as in ridge waveguide QCLs as in buried heterostructure QCLs. Although the role of spatial hole burning (SHB) in lowering the second threshold was understood, it was thought [16] that the low-threshold RNGH instability does not occur just as a result of the induced grating of carrier population. Therefore, an additional assumption was made in [16] on a built-in saturable absorber in the cavity of QCL, allowing for a reduction of the 2nd threshold from 9-fold excess to about 1.1 times above the lasing threshold. As a matter of fact, the saturable absorber has been shown to lower the instability threshold in a CW ring laser [17]. However, the nature of saturable absorption in QCLs has never been fully clarified [1]. In particular, the authors in [16] have evoked the Kerr lensing effect as a possible mechanism responsible for the saturable absorption. In case of narrow ridge waveguide lasers, due to overlap of the waveguide mode tails and the metal contact deposited on the waveguide, the Kerr lensing does may lead to saturable absorption. However it cannot produce the saturable absorption effect of the same strength in buried heterostructure QCLs. At the same time low-threshold RNGH instability was observed in both ridge waveguide and buried heterostructure QCLs.

In this paper, we develop a model for low-threshold multimode RNGH instability in a QCL without saturable absorber. This is a more realistic representation for a monolithic single-section *FP cavity laser*. Technically, a set of important corrections to the original treatment [7,8,16] was made. It consists in (i) accounting for the induced carrier coherence grating alongside with the carrier population grating and in (ii) accounting for the carrier diffusion process that leads to relaxation of both gratings. The outcomes of our analysis for the second threshold in QCLs are in perfect agreement with the numerical simulations based on a travelling wave (TW) rate equation model as well as with the experimental data available in the literature [6,16,18]. Interestingly, we find that regular self-pulsations at picosecond time scale are possible in short-cavity QCLs. Our model also explains the nonexistence of multimode RNGH instability in quantum well (QW) laser diodes (LDs).

This paper is organized in the following manner: The model is presented in Sec. II. Section III addresses the second threshold conditions and discusses the RNGH instability in long-cavity and short-cavity devices, as well as the importance of the carrier diffusion effects in QCLs and conventional LDs; Sec. IV provides a conclusion.

## II. MODEL DESCRIPTION

### A. QCL laser as a model system for RNGH instabilities

A quantum cascade laser is a unipolar device based on intersubband transitions. Its active region consists of a large number of periodically repeated epilayers. Each period (stage) comprises a gain and injection/relaxation regions. While the gain region serves to create a population inversion between the two levels of the lasing transition, the purpose of the injection/relaxation region is to depopulate the lower lasing level and to provide injection of electrons in the upper lasing level of the next stage [1]. Although QCLs have a complicated structure of epilayers, it was understood and confirmed experimentally that the internal structure does not alter directly the appearance of multimode RNGH instabilities and that a semiclassical model with effective parameters of the laser gain medium suffices [16]. Alongside with the studies of RNGH instabilities in InGaAs QCLs, we apply our analytic model to QW LDs based on AlGaAs and InGaN alloys. For all of them we assume a simple single-section Fabry-Perot (FP) cavity design. The parameters of our model for three types of lasers considered here are summarized in Table I.

**TABLE I. Dynamic model parameters for QCLs and QW LDs considered in this paper.**

| Parameter | Name | InGaAs QCL | AlGaAs QW LD | InGaN QW LD |
|---|---|---|---|---|
| $\Lambda$ | Lasing wavelength | 10 μm | 850 nm | 420 nm |
| $T_1$ | Carrier lifetime | 1.3 ps | 1-2 ns | |
| $T_2$ | Carrier dephasing time | 140 fs | 100 fs | |
| $T_{2_eff}$ | Effective carrier dephasing time in the presence of diffusion | 138 fs | 86 fs | 84 fs |
| $T_g$ | Relaxation time of the carrier population grating | 0.927 ps | 0.158 ps | 0.13 ps |
| $T_{2_g}$ | Relaxation time of the coherence grating | 128 fs | 41 fs | 37 fs |
| $\alpha_i$ | Intrinsic material loss | 24 $cm^{-1}$ | 5 $cm^{-1}$ | 35 $cm^{-1}$ |
| D | Diffusion coefficient | 180 $cm^2/s$ | 20 $cm^2/s$ (ambipolar) | 7 $cm^2/s$ (ambipolar) |
| $n_g$ | Group refractive index | 3.3 | 3.8 | 3.5 |
| $R_1$, $R_2$ | Cavity facet reflection coefficients | 27% | 27% | 18 % |
| $\Gamma$ | Optical mode confinement factor | 0.5 | 0.01 | 0.02 |
| $\partial g/\partial n$ | Differential material gain | $2.1\times10^{-4}$ $cm^3/s$ | $1\times10^{-6}$ $cm^3/s$ | $2.2\times10^{-6}$ $cm^3/s$ |
| $n_t$ | Transparency carrier density | $7\times10^{14}$ $cm^{-3}$ | $2\times10^{18}$ $cm^{-3}$ | $1.6\times10^{19}$ $cm^{-3}$ |

As a model system for RNGH instability, we consider an InGaAs QCL with a direct transition, having the carrier lifetime $T_1$ of about one picosecond. This is much smaller than the cavity roundtrip time. As in the case of heterostructure epilayers, we do not examine any particular lateral cavity design since it was shown experimentally that the laser design (ridge waveguide [16] or buried heterostructure [6]) has no or little impact on the RNGH instability in QCLs. Although our QCL model parameters are slightly different from the ones in Ref.[16], this difference has no impact on the main conclusions of the present paper, in particular, regarding the occurrence of low-threshold RNGH instability.  In the table, the optical mode confinement factor and transparency carrier density for QCL are defined using the overall thickness of epilayers in the period.

In Section III.E we apply our model to AlGaAs QW LDs and InGaN QW LDs (last two columns in Table I) with the objective to verify the model agreement with experimental observations that single-section QW LDs do not reveal any feature of RNGH instability.

For all considered here model systems, the photon lifetime in the cavity exceeds the dephasing time $T_2$, so as the dynamic behavior reported here is intrinsically different from the Class-D laser dynamics discussed in [19].

### B. Travelling wave rate equation model

As a starting point of our analysis we use semiclassical Maxwell-Bloch (MB) equations for a two-level system. Following along the lines of the travelling wave model approach from [20] and accounting for the diffusion term in the Schrödinger equation [21], we obtain the following system of rate equations:

$$\dot{\rho}_{ab} = i\omega\rho_{ab} + i\frac{\mu E}{\hbar}\Delta - \frac{\rho_{ab}}{T_2} + D\frac{\partial^2 \rho_{ab}}{\partial z^2} \tag{1}$$

$$\dot{\Delta} = -2i\frac{\mu E}{\hbar}(\rho^*_{ab} - \rho_{ab}) + \frac{\Delta_{pump} - \Delta}{T_1} + D\frac{\partial^2 \Delta}{\partial z^2} \tag{2}$$

$$\partial^2_z E - \frac{n_g^2}{c^2}\partial^2_t E = \Gamma\frac{N\mu}{\varepsilon_0 c^2}\partial^2_t(\rho^*_{ab} + \rho_{ab}) + \frac{n_g l'_0}{c}\partial_t E \tag{3}$$

where $\rho_{ab} = \rho_{ba}^{*}$ is the off-diagonal element of the density matrix, the non-equilibrium carrier density $N\Delta = N(\rho_{bb} - \rho_{aa})$ is represented as the population inversion between the upper (index "*bb*") and lower (index "*aa*") levels of the lasing transition, $N\mu(\rho_{ab}^{*} + \rho_{ab})$ is the active medium polarization ("coherence"), $\omega$ and $\mu$ denote the resonant frequency and the dipole matrix element of the lasing transition, respectively, $T_1$ and $T_2$ are the longitudinal and transverse relaxation times, $N\Delta_{pump} / T_1$ is the pump rate due to injection of electrons into the upper lasing level, $N$ is defined by the doping density in the injector region and $D$ is the diffusion coefficient for electrons in the plane of active QWs of the QCL structure (for a LD, the ambipolar diffusion coefficient for the holes and electrons is used instead). $E$, $\Gamma$ and $n_g$ stand for the cavity mode electric field, the overlap between the optical mode and the active region and the group refractive index for the cavity mode, respectively. Note that the differential material gain is $\partial g / \partial n = \omega T_2 \mu^2 / \hbar n_g^2 \varepsilon_0$ and $l_0$ is the linear loss coefficient. The difference between our MB equations for QCLs and equations used in [16] is the coherence diffusion term $D\partial^2 \rho_{ab} / \partial z^2$ that appears in the first equation. Numerical simulations with the travelling wave rate equation model reported below were accomplished by introducing in (1)-(3) two slowly varying amplitudes for the counter propagating waves in the Fabry-Perot (FP) cavity and distinguishing the medium's polarizations associated with the forward and backward traveling waves [22].

### C. Approximation of coupled-mode rate equation model

The system of MB equations (1)-(3) is not ideally suited for analytical study on multimode RNGH instability. In Ref [23] a comprehensive approach for analysis of multimode instabilities in a ring laser was established using a truncated set of coupled-mode equations. A similar approach was applied in [16] for the case of FP-cavity QCLs. However the key point of such truncated modal expansion is to take into account all coupled harmonics *in a self-consistent way* [23]. We argue that previous theoretical treatment of RNGH instabilities in QCLs [16] does not take into account the spatial harmonics of induced macroscopic polarization of the gain medium *in the FP-cavity laser or bidirectional ring laser*. As a result, in the limit of adiabatic elimination of the medium polarization, the truncated set of coupled-mode equations from [16] does not agree with well-established models for a Class-B laser dynamics [24]. (See discussion to equations (7)-(11) below and also the Appendix A. We stress that by no means our considerations for *bidirectional ring lasers* and *FP cavity lasers* can be applied to the *unidirectional ring laser* considered in Refs. [7,8]). Furthermore, in order to match the experimentally measured second threshold of ~$1.1 \cdot I_{\text{th}}$ ($I_{\text{th}}$ is the lasing threshold pump current), the model in [16] assumes that there is a built-in saturable absorber in the monolithic cavity of QCL due to the Kerr lensing effect. As a matter of fact, the saturable absorber does lower the RNGH instability threshold in a CW laser [17]. However in the buried heterostructure QCLs, which also exhibit low second threshold, the proposed Kerr-lensing mechanism for saturable absorption has never been unambiguously confirmed [1].

In order to elucidate another possible origin of the low-threshold RNGH instability, we have expanded the MB equations (1)-(3) in to a truncated set of self-consistent coupled mode equations. Our analysis of the SHB effect has shown that as soon as the spatial grating of carrier population is taken into account (the terms $\Delta_2^{\pm}$ below), the third-order spatial harmonics of the induced macroscopic polarization grating (the terms $\eta_{++}$ and $\eta_{--}$ below) have to be accounted for in the coupled-mode expansion:

$$E(z,t) = \frac{1}{2}\left[E_+^* e^{-i(\omega t - kz)} + E_+ e^{i(\omega t - kz)}\right] + \frac{1}{2}\left[E_-^* e^{-i(\omega t + kz)} + E_- e^{i(\omega t + kz)}\right] \tag{4}$$

$$\rho_{ab}(z,t) = \left(\eta_+ + \eta_{++} e^{-2ikz}\right) e^{i(\omega t - kz)} + \left(\eta_- + \eta_{--} e^{2ikz}\right) e^{i(\omega t + kz)} \tag{5}$$

$$\Delta(z,t) = \Delta_0 + \Delta_2^+ e^{2ikz} + \Delta_2^- e^{-2ikz}, \tag{6}$$

where $E(z,t)$ and $\Delta(z,t)$ are real-valued variables ($\Delta_2^{+*} = \Delta_2^-$) and the amplitudes $E_\pm$, $\eta_\pm$, $\eta_{\pm\pm}$, $\Delta_0$ and $\Delta_2^\pm$ vary slowly in time and space as compared to the plane wave carriers. The two optical carrier waves propagating at frequency ω and wavenumbers ±*k* stand for the initial lasing mode in the cavity.

Substitution of Eqs. (4)-(6) in MB equations (1)-(3) leads to the following coupled-mode equations in the slowly varying envelope approximation:

$$\frac{n}{c}\partial_t E_\pm = \mp\partial_z E_\pm - i\frac{N\mu\Gamma\omega}{cn_g\varepsilon_0}\eta_\pm - \frac{1}{2}l_0 E_\pm, \tag{7}$$

$$\partial_t \eta_\pm = \frac{i\mu}{2\hbar}\left(\Delta_0 E_\pm + \Delta_2^\mp E_\mp\right) - \frac{\eta_\pm}{T_2} - k^2 D\eta_\pm, \tag{8}$$

$$\partial_t \eta_{\pm\pm} = \frac{i\mu}{2\hbar}E_\pm\Delta_2^\mp - \frac{\eta_{\pm\pm}}{T_2} - 9Dk^2\eta_{\pm\pm}, \tag{9}$$

$$\partial_t \Delta_0 = \frac{\Delta_{pump} - \Delta_0}{T_1} + \frac{i\mu}{\hbar}\left(E_+^*\eta_+ + E_-^*\eta_- - c.c\right), \tag{10}$$

$$\partial_t \Delta_2^\pm = \frac{i\mu}{\hbar}\left(E_\pm^*\eta_\mp - E_\mp\eta_\pm^* - E_\pm\eta_{\pm\pm}^* + E_\mp^*\eta_{\mp\mp}\right) - \frac{\Delta_2^\pm}{T_1} - 4k^2 D\Delta_2^\pm \tag{11}$$

where $l_0$ is the cavity loss coefficient that comprises intrinsic material losses and output coupling losses. In order to verify our coupled-mode expansion (4)-(6), we have performed the adiabatic-following approximation test [9] (see Appendix A) and find that the adiabatic approximation for our set of equations is in excellent agreement with the well-established Class-B laser model. Note that this is not the case for a set of coupled-mode equations used in [16]. The second important difference is that our coupled-mode system (7)-(11) shows high-frequency instabilities at Rabi oscillation frequency while the one of Ref.[16] shows instabilities at much lower frequency (see Section III.F).

Next, we carry out a linear stability analysis of our model system (7)-(11). In what follows we introduce the effective relaxation times that account for the contribution from the diffusion terms

$T_g = \left(T_1^{-1} + 4Dk^2\right)^{-1}$, $T_{2_g} = \left(T_2^{-1} + 9Dk^2\right)^{-1}$ and $T_{2_eff} = \left(T_2^{-1} + Dk^2\right)^{-1}$. (These relaxation times are quoted in Table 1.) We also define new variables $e_\pm = l_0 E_\pm \mu / \hbar$, $\pi_\pm = \eta_\pm l_0 / (\Delta_{th} T_{2_eff})$, $n_0 = \Delta_0 l_0 / (\Delta_{th} T_{2_eff})$, $n_\pm = \Delta_\pm l_0 / (\Delta_{th} T_{2_eff})$. The pump rate is accounted for by the parameter $p = \Delta_{pump} / \Delta_{th}$, which measures the pump rate excess above the lasing threshold in the absence of SHB, and $N\Delta_{th} = \hbar \varepsilon_0 n_g l_0 \big/ \Gamma \omega T_{2_eff} \mu^2$ is the carrier density at the lasing threshold.

We obtain the steady-state solution of Eqs.(7)-(11) assuming an arbitrary initial phase $\theta$ of the wave propagating in the positive z axis direction (forward wave). Because the optical mode field $E(z,t)$ and the population parameter $\Delta(z,t)$ in Eqs.(4)-(6) are the real-valued variables, the phase of the wave propagating in the backward direction is $-\theta$, yielding the following solution for CW lasing regime:

$$e_\pm = e^{\pm i\theta} l_0 \mathcal{E},$$

$$n_0 = \frac{\nu_0 l_0}{T_{2_eff}}, \quad n_\pm = \frac{-e^{\mp i2\theta} l_0 (\nu_0 - 1)}{T_{2_eff}}, \tag{12}$$

$$\pi_\pm = \frac{i}{2} e^{\pm i\theta} l_0 \mathcal{E}, \quad \pi_{\pm\pm} = -\frac{T_{2_g}}{T_{2_eff}} \frac{i}{2} (\nu_0 - 1) e^{\pm i3\theta} l_0 \mathcal{E},$$

where

$$\mathcal{E} = \sqrt{\frac{p - \nu_0}{2T_1 T_{2_eff}}} \tag{13}$$

is the normalized field amplitude. Note that variables $n_-^* = n_+$, $e_-^* = e_+$, are complex conjugate while the variables $\pi_-^* = -\pi_+$ , $\pi_{--}^* = -\pi_{++}$ are anti-conjugate.

The SHB increases the effective lasing threshold and reduces the slope efficiency. To account for this effect, we introduce an additional parameter $\nu_0 = \Delta_0 / \Delta_{th}$, which is the ratio of the fundamental harmonic of the average carrier density $N\Delta_0$ to its value at the lasing threshold $N\Delta_{th}$ (at $p$=1), when SHB effect does not yet settle–in. We obtain the following expression for $\nu_0$:

$$\nu_0=\frac{1}{2}\left(p+1+\frac{T_{2_eff}}{T_{2_g}}+\frac{2T_1T_{2_eff}}{T_gT_{2_g}}\right)-\sqrt{\frac{1}{4}\left(p+1+\frac{T_{2_eff}}{T_{2_g}}+\frac{2T_1T_{2_eff}}{T_gT_{2_g}}\right)^2-p\left(1+\frac{T_{2_eff}}{T_{2_g}}\right)-\frac{2T_1T_{2_eff}}{T_gT_{2_g}}}\ . \quad (14)$$

At lasing threshold, $\nu_0=1$ and it increases above the threshold (at $p$>1). At very high pump rate ($p$>>1), $\nu_0$ asymptotically approaches the value of $1+T_{2_eff}/T_{2_g}$ .

The linear stability analysis of the steady-state solution (12)- (14), is performed by introducing small perturbations $\delta e_+=\delta e_r+i\delta e_i$, , $\delta\pi_+=\delta\pi_r+i\delta\pi_i$ , $\delta\pi_{++}=-(\delta\pi_{rr}+i\delta\pi_{ii})$, $\delta n_0$ and $\delta n_+=\delta n_r+i\delta n_i$ in the corresponding variables of the coupled-mode equations (7)-(11). Taking into account the complex conjugate and anti-conjugate relationships between the variables $\delta n_-=\delta n_+^*$, $\delta e_-=\delta e_+^*$, $\delta\pi_-=-\delta\pi_+^*$, $\delta\pi_{--}=-\delta\pi_{++}^*$, we obtain the following linearized system of differential equations:

$$\frac{d}{dt}\begin{bmatrix}\delta\pi_i\\ \delta e_r\\ \delta n_0\\ \delta n_r\\ \delta\pi_{ii}\\ \delta\pi_r\\ \delta e_i\\ \delta n_i\\ \delta\pi_{rr}\end{bmatrix}=\begin{bmatrix}
-\frac{1}{T_{2_eff}} & \frac{\nu_0-(\nu_0-1)\cos 2\theta}{2T_{2_eff}} & \frac{\varepsilon\cos\theta}{2} & \frac{\varepsilon\cos\theta}{2} & 0 & 0 & -\frac{(\nu_0-1)\sin 2\theta}{2T_{2_eff}} & -\frac{\varepsilon\sin\theta}{2} & 0\\
\frac{1}{\tau} & -\frac{1}{2\tau}-i\Omega & 0 & 0 & 0 & 0 & 0 & 0 & 0\\
-4\varepsilon\cos\theta & -2\varepsilon\cos\theta & -\frac{1}{T_1} & 0 & 0 & 4\varepsilon\sin\theta & -2\varepsilon\sin\theta & 0 & 0\\
-2\varepsilon\cos\theta & \left[(\nu_0-1)\frac{T_{2_g}}{T_{2_eff}}\cos 3\theta-\cos\theta\right]\varepsilon & 0 & -\frac{1}{T_g} & 2\varepsilon\cos\theta & -2\varepsilon\sin\theta & \left[(\nu_0-1)\frac{T_{2_g}}{T_{2_eff}}\sin 3\theta+\sin\theta\right]\varepsilon & 0 & -2\varepsilon\sin\theta\\
0 & \frac{(\nu_0-1)\cos 2\theta}{2T_{2_eff}} & 0 & -\frac{\varepsilon\cos\theta}{2} & -\frac{1}{T_{2_g}} & 0 & -\frac{(\nu_0-1)\sin 2\theta}{2T_{2_eff}} & -\frac{\varepsilon\sin\theta}{2} & 0\\
0 & \frac{(\nu_0-1)\sin 2\theta}{2T_{2_eff}} & -\frac{\varepsilon\sin\theta}{2} & \frac{\varepsilon\sin\theta}{2} & 0 & -\frac{1}{T_{2_eff}} & -\frac{\nu_0+(\nu_0-1)\cos 2\theta}{2T_{2_eff}} & \frac{\varepsilon\cos\theta}{2} & 0\\
0 & 0 & 0 & 0 & 0 & -\frac{1}{\tau} & -\frac{1}{2\tau}-i\Omega & 0 & 0\\
2\varepsilon\sin\theta & \left[-(\nu_0-1)\frac{T_{2_g}}{T_{2_eff}}\sin 3\theta+\sin\theta\right]\varepsilon & 0 & 0 & 2\varepsilon\sin\theta & -2\varepsilon\cos\theta & \left[(\nu_0-1)\frac{T_{2_g}}{T_{2_eff}}\cos 3\theta+\cos\theta\right]\varepsilon & -\frac{1}{T_g} & 2\varepsilon\cos\theta\\
0 & -\frac{(\nu_0-1)\sin 2\theta}{2T_{2_eff}} & 0 & \frac{\varepsilon\sin\theta}{2} & 0 & 0 & -\frac{(\nu_0-1)\cos 2\theta}{2T_{2_eff}} & -\frac{\varepsilon\cos\theta}{2} & -\frac{1}{T_{2_g}}
\end{bmatrix}\cdot\begin{bmatrix}\delta\pi_i\\ \delta e_r\\ \delta n_0\\ \delta n_r\\ \delta\pi_{ii}\\ \delta\pi_r\\ \delta e_i\\ \delta n_i\\ \delta\pi_{rr}\end{bmatrix}$$

(15)

where $\tau=n_g/cl_0$ is the photon lifetime in the cavity.

Using the ansatz from Ref. [8], we recast perturbations to the main lasing mode in the form of propagating waves $\propto \exp(in_g \Omega z / c + \Lambda t)$, where Λ can be interpreted as Lyapunov exponent, $\Omega n_g$/c is the detuning of the propagation constant from the one of the lasing mode and $-\mathrm{Im}(\Lambda)$ is the frequency offset. Note that among all possible solutions Λ(Ω) of the dispersion equation that follows from Eq.(15), only that solutions with positive increment Re(Λ)>0 will build-up in the cavity after several roundtrips, which fulfill the cavity roundtrip phase self-repetition condition $2\Omega L n_g/c=2\pi n$, where $L$ is the cavity length and $n$ is an integer number. Nevertheless, it is convenient to carry out our stability analysis considering $\Omega$ as a continuous parameter. It can be regarded as the offset frequency from the lasing mode because for all practical situations of interest in this study, the approximation $\mathrm{Im}(\Lambda) \approx -\Omega$ suffices.

Numerical solution of Eq.(15) reveals that only one Lyapunov exponent of the 9×9 linear stability matrix may have positive real part and thus may lead to a multimode instability in the practical range of pump rates considered here (see Fig. 1 below and related discussion).

Via a transformation of variables, the 9×9 matrix in Eq.(15) can always be altered into a block-diagonal form with 5×5 and 4×4 blocks, as depicted in Eq.(16) below. Out of these two matrices, only the smaller-size matrix may exhibit Lyapunov exponent with the positive real part in the practically feasible range of pump currents (see discussion to Fig.1 below). The required transformation and hence the set of variables responsible for instability are both changing with the initial phase of the mode $\theta$. However the instability increment Re(Λ) remains fixed, attesting that the initial phase has no impact on the occurrence of instability. For simplicity, in what follows we assume that initially the laser operates in a CW lasing regime and the phase of the mode is $\theta$=0, yielding us the following eigenproblem:

$$
\begin{bmatrix}
-\frac{1}{T_{2_eff}}-\Lambda & \frac{1}{2T_{2_eff}} & \frac{\mathcal{E}}{2} & \frac{\mathcal{E}}{2} & 0 & 0 & 0 & 0 & 0 \\
\frac{1}{\tau} & -\frac{1}{2\tau}-i\Omega-\Lambda & 0 & 0 & 0 & 0 & 0 & 0 & 0 \\
-4\mathcal{E} & -2\mathcal{E} & -\frac{1}{T_1}-\Lambda & 0 & 0 & 0 & 0 & 0 & 0 \\
-2\mathcal{E} & \left[(\nu_0-1)\frac{T_{2_g}}{T_{2_eff}}-1\right]\mathcal{E} & 0 & -\frac{1}{T_g}-\Lambda & 2\mathcal{E} & 0 & 0 & 0 & 0 \\
0 & \frac{(\nu_0-1)}{2T_{2_eff}} & 0 & -\frac{\mathcal{E}}{2} & -\frac{1}{T_{2_g}}-\Lambda & 0 & 0 & 0 & 0 \\
0 & 0 & 0 & 0 & 0 & -\frac{1}{T_{2_eff}}-\Lambda & -\frac{(2\nu_0-1)}{2T_{2_eff}} & \frac{\mathcal{E}}{2} & 0 \\
0 & 0 & 0 & 0 & 0 & -\frac{1}{\tau} & -\frac{1}{2\tau}-i\Omega-\Lambda & 0 & 0 \\
0 & 0 & 0 & 0 & 0 & -2\mathcal{E} & \left[(\nu_0-1)\frac{T_{2_g}}{T_{2_eff}}+1\right]\mathcal{E} & -\frac{1}{T_g}-\Lambda & 2\mathcal{E} \\
0 & 0 & 0 & 0 & 0 & 0 & -\frac{(\nu_0-1)}{2T_{2_eff}} & -\frac{\mathcal{E}}{2} & -\frac{1}{T_{2_g}}-\Lambda
\end{bmatrix}
\cdot
\begin{bmatrix}
\delta\pi_i \\ \delta e_r \\ \delta n_0 \\ \delta n_r \\ \delta\pi_{ii} \\ \delta\pi_r \\ \delta e_i \\ \delta n_i \\ \delta\pi_{rr}
\end{bmatrix}
=0 \qquad (16)
$$

We find that only one eigenvalue of the lower 4×4 block may have Re(Λ)>0 and may lead to unstable CW lasing regime observed in experiment. The upper 5×5 block shows only stable solutions in a large range of pump rates above the lasing threshold. Thus for a QCL from Table 1, this behavior is observed up to *p*~60 times above the lasing threshold (Fig. 1). For an AlGaAs QW LD from Table 1, it continuous up to *p*~1100. In what follows we limit discussion to this sufficiently large range of pump rates when only one eigenvalue in Eq. (16) may lead to a multimode instability.

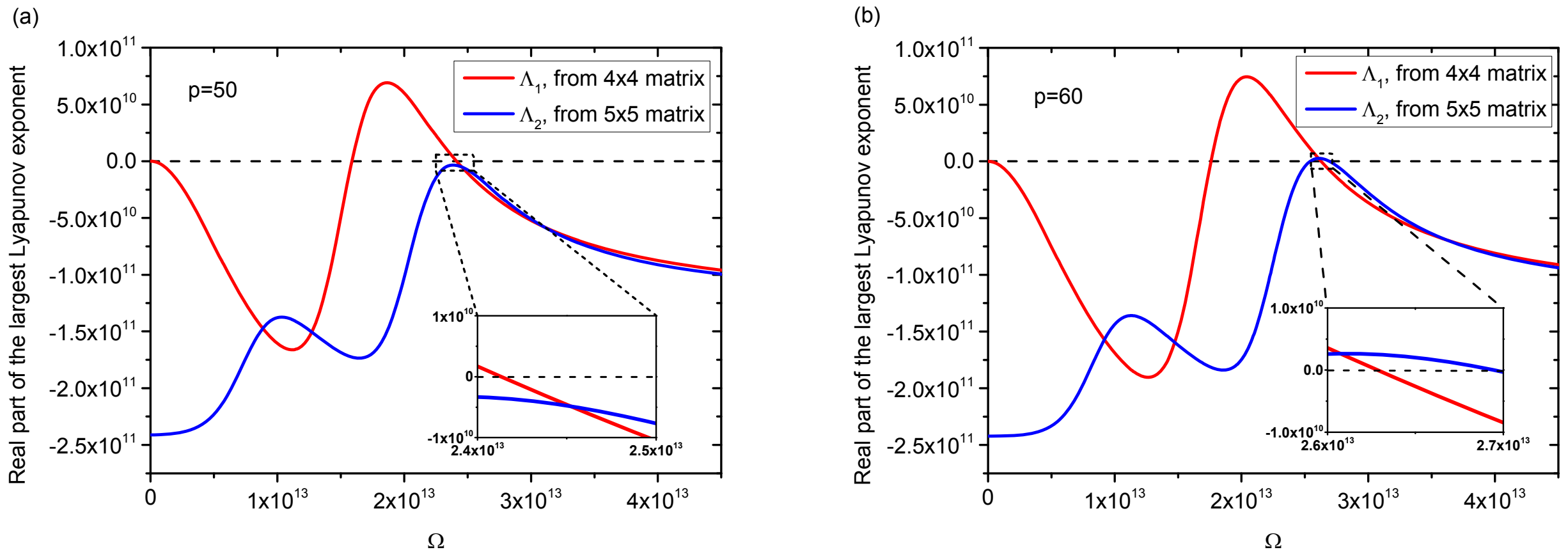

FIG. 1. Spectra of instability increments (largest real parts of the Lyapunov exponents) of the 4×4 (red curves) and 5×5 matrix (blue curves) blocks at *p*=50 (panel (a)) and *p*=60 (panel (b)) for QCL with parameters from Table 1.

The instability in Eq.(16) is related to variables $\delta\pi_r$, $\delta e_i$, $\delta n_i$ and $\delta\pi_{rr}$. When the initial phase of the mode $\theta$ is detuned from $\theta$=0, the variables associated with the lower 4×4 block are altered. For example, for $\theta = \pi / 2$, the instability is due to perturbations $\delta\pi_i$, $\delta e_r$, $\delta n_r$ and $\delta\pi_{ii}$. Therefore, it is not possible to identify the nature of instability (RNGH-like or of a different kind) from the set of variables involved. At the same time we find that whatever the initial phase $\theta$, the offset frequency $\Omega_{max}$ at the maximum increment Re(Λ) is close to the Rabi oscillation frequency $\Omega_{Rabi} = \sqrt{(p - v_0)/T_1 T_{2_eff}}$ (see Section III below). This allows us to attribute this instability to the RNGH-like behavior discussed in Refs.[7,8].

This behavior should be contrasted with the one reported in Ref.[16], where (i) the carrier coherence grating (9) has not been taken into account and (ii) truncated coupled mode equations for carrier population grating and medium polarization contained errors (see our discussion to Eqs. (7)-(11), Appendix A and our analytical solution obtained in Ref. [25]). If one omits the variables $\delta\pi_{rr}$ and $\delta\pi_{ii}$ associated with the carrier coherence grating, the lower 4×4 matrix block in Eq.(16) simply becomes of 3×3 size. It still exhibits an instability at small pump rate above the lasing threshold. According to Ref. [16] the instability occurs at significantly lower frequencies around $\Omega_{SHB} = \sqrt{T_1^{-1}\sqrt{(p-1)/3T_1T_2}} \sim \sqrt{\Omega_{Rabi}/T_1} / \sqrt[4]{3}$. Therefore this instability has been attributed in [16] to the SHB effect. Below we show that this instability occurs in fact at much higher frequency close to $\Omega_{Rabi}$. These outcomes of our numerical studies agree very well with our analytic solution reported in Ref. [25].

The upper 5×5 matrix block in (16), which shows only stable solutions in the presence of the carrier coherence grating (9) and in the reasonable range of pump rates, takes the form of a 4×4 matrix after dropping out the variables $\delta\pi_{rr}$ and $\delta\pi_{ii}$. Our analysis shows that it may lead to RNGH instability at the pump rate approximately 10 times above the lasing threshold, in agreement with the theoretical predictions from [7,8]. The second threshold always remains that high unless a saturable absorber is introduced to the model system [17], the possibility which was studied in [16].

We argue that inclusion of both grating terms (9) and (11) allows one to obtain low-threshold RNGH instability in a *FP cavity laser* (or *bidirectional ring laser*) without saturable absorber. This instability is of different origin from the one discussed in Ref.[16]. In our case it originates from the lower 4×4 matrix block in Eq.(16) and not from the upper 5×5 block. The coherence grating (9) renders the second threshold associated with the 5×5 block further more prohibitively higher (at p~60 instead of the initial value of p~9-10 in the original RNGH instability case for a *unidirectional ring laser*). In the next section we elucidate these points in further details.

## III. RESULTS AND DISCUSSION

### A. RNGH instability threshold

For QCLs that have the cavity length of 2-4 mm, our linear stability analysis predicts the multimode RNGH instability at a pump rate of a few percent above the lasing threshold. This prediction is in agreement with numerous experimental data available in the literature [6,16,18]. Our numerical simulations based on TW rate equation model also confirm this behavior and the transition to multimode operation at just above the lasing threshold (see Section III.C).

In the case of a short-cavity QCL of just 100 μm long, our numerical simulations reveal stable CW operation regime up to the pump rate of 2-2.5 times above the lasing threshold. The RNGH-like self-pulsations occur above the second threshold of $p_{th2}$~2.5 (see below). However, the linear stability analysis based on Eq.(16) shows that the real part of the Lyapunov exponent Re(Λ) becomes positive at the pump rate $p$ just slightly above 1. Moreover, with increasing $p$, the instability increment Re(Λ) monotonically increases without revealing any specific behavior at $p_{th2}$. Therefore we examine the spectral behavior of the increment Re(Λ) and we arrive to a new insight into the second threshold condition, which we introduce below.

Figure 2 illustrates the spectral behavior of the increment Re(Λ) in a QCL with the cavity length of 100 μm. Only positive offset frequencies are shown, because the spectral shape of Re(Λ) is an even function of Ω. The instability increment is plotted being normalized to the cavity mode separation, which yields us the spectrum of the roundtrip gain coefficient $2\mathrm{Re}(\Lambda)n_gL/c$ for various cavity modes. In the figure, the instability gain spectra are depicted at three different pump rates of $p$=1.1, 1.2 and 3 (green, red and blue curves, respectively). We also indicate the locations of a few cavity modes by plotting the Airy function of the "cold cavity" (black curve, not in scale). The zero offset frequency corresponds to the initially lasing mode.

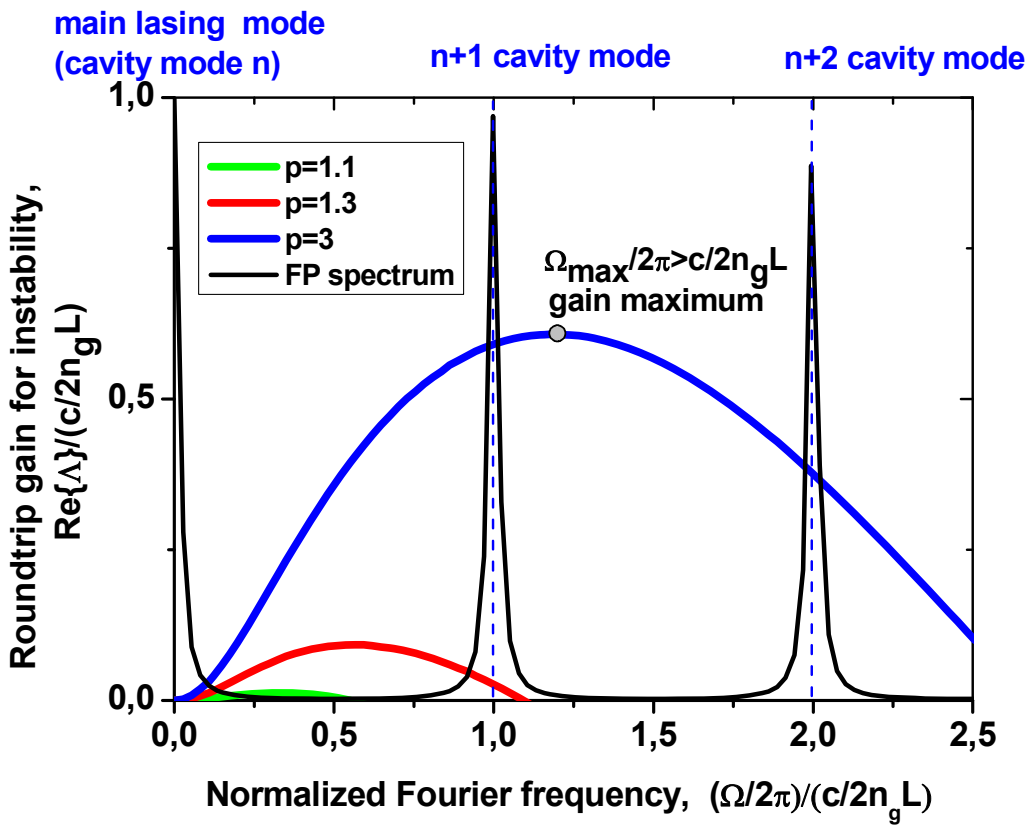

FIG. 2. Positive offset frequency part of the spectrum of the roundtrip gain for instability in a 100 μm long QCL at normalized pump rate of $p$=1.1 (green curve), $p$=1.3 (red curve) and $p$=3. The black curve indicates the Airy function of the cold cavity (not in scale). Other QCL parameters are given in Table I.

The RNGH condition [7,8] for multimode instability in a CW operating single-mode laser requires that the instability increment is positive at a frequency of another cavity mode. An example that matches this definition is the instability gain curve plotted in Fig. 2 for $p$=1.3 (red curve). It shows a positive gain for the first two adjacent cavity modes at the offset frequencies $\Omega/2\pi=\pm c/2n_gL$. In Fig. 3, this gain coefficient is plotted as a function of the pump rate (dashed blue curve). According to the conventional definition of the second threshold, the multimode RNGH instability should occur at $p$=1.25, when the gain for instability becomes positive at some of the non-lasing cavity modes.

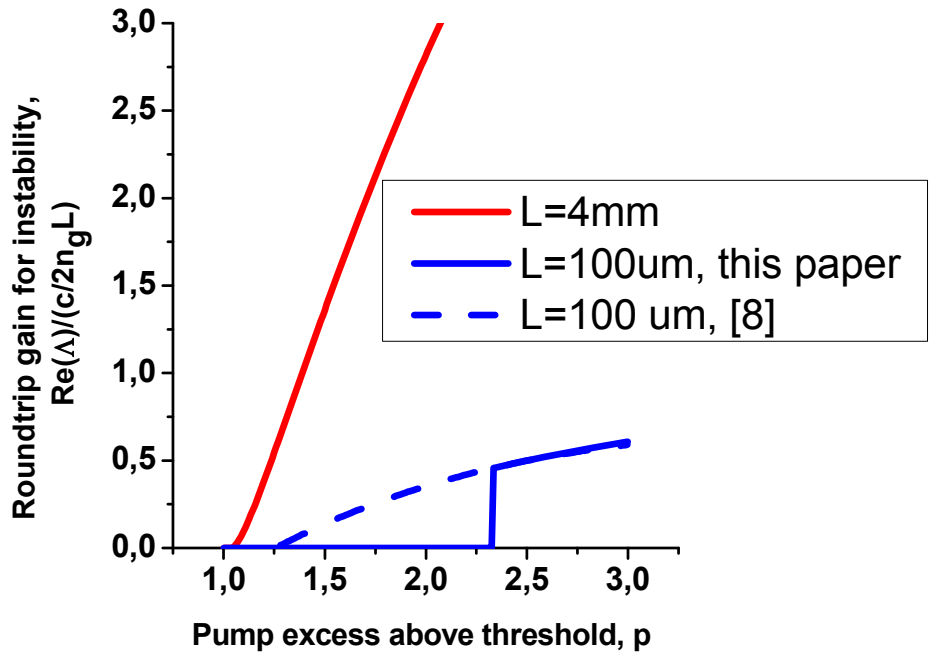

FIG. 3. Lyapunov stability analysis: roundtrip gain for RNGH instability vs. pump excess above threshold for the QCLs with the cavity lengths of 4 mm (red curve) and 100 μm (solid blue curve). For explanations on the dashed blue curve, see discussion in the text. In long-cavity QCL, the second threshold is very low, at $p_{th2}$=1.05, while $p_{th2}$=2.35 in the short-cavity QCL. Other QCL parameters are given in Table I.

In order to verify the second threshold condition, we perform a series of numerical simulations based on the travelling wave rate equation model [22]. As a model system we use a short-cavity single-section sample (L=100 μm). The travelling wave model incorporates Langevin force terms that seed the spontaneous polarization noise into the system (these terms are not indicated in Eqs. (1)-(3), but an example can be found in [22]). We examine several different power levels of the noise injected into the system. Fig. 4 shows that the output power of amplified spontaneous emission $P_{sp}$ varies over 6 orders of magnitude for the range of spontaneous polarization noise used here.

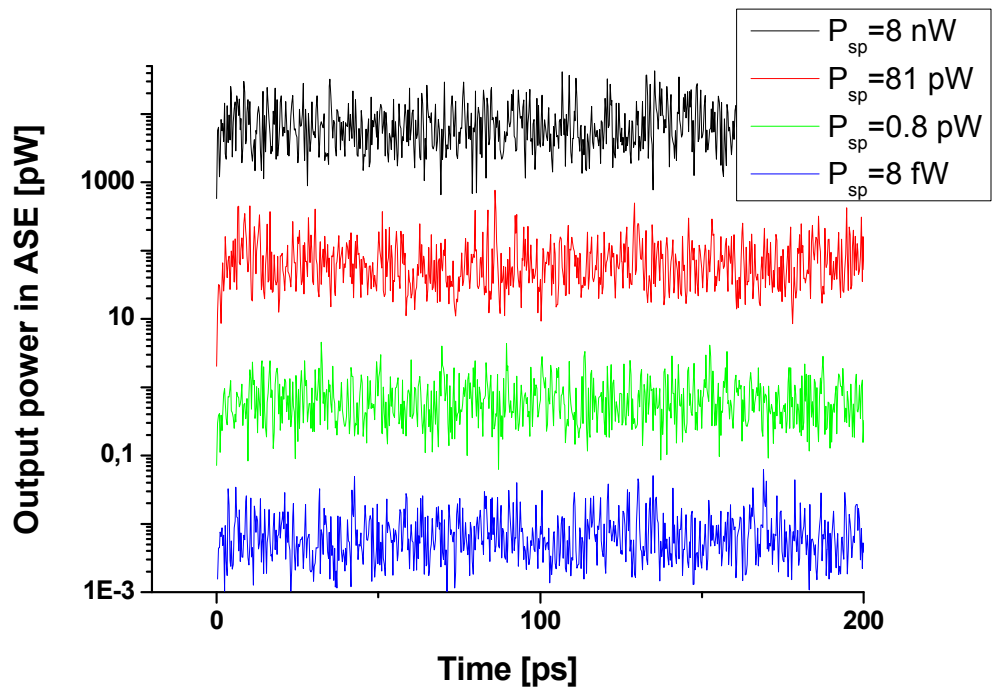

FIG. 4. Output power variation in amplified spontaneous emission (ASE) regime below the lasing threshold ($p$=0.9) for different levels of spontaneous polarization noise used in the numerical simulations. QCL cavity length is 100 μm. Other parameters are given in Table I.

In Fig 5, for each set of parameters ($p$, $P_{sp}$) we perform a series of 20 simulations and calculate the probability of occurrence of the RNGH like self-pulsations (various examples of the waveforms can be found in Figs. 9 (c)-(f) below). These statistical data are displayed in Fig. 5 as vertical bars with the color corresponding to the spontaneous polarization noise power in Fig.4. It follows that there are no systematic correlations between the occurrence of RNGH instability and spontaneous noise power $P_{sp}$ injected into the system. In Fig 5 we also plot the average probability over all 80 realizations at different $P_{sp}$ (solid curve). It has the dispersion of ±0.056, which is indicated in Fig. 5 by the error bars.

The data in Fig. 5 attest that there is no multimode RNGH instability at the pump rate $p$=1.5. This is in contradiction with the conventional definition of the second threshold, foreseeing $p_{th2}$=1.25 (Fig 3, dashed blue curve). At $p$=2, only a few realizations have resulted in RNGH self-pulsations. The average probability of these is of 0.04, which is below the uncertainty limit. The RNGH instability develops at $p$ between 2 and 2.5, with p=2.5 being the first point in Fig. 5 for which the probability exceeds the uncertainty range.

Obviously, the results of our numerical simulations are not in agreement with the second threshold condition as originally proposed by Risken and Nummedal [7], and Graham and Haken [8]. Analyzing the data in Fig.5, we find that RNGH instability occurs when the instability gain maxima located at the offset frequencies $\Omega_{\max} \approx \pm\Omega_{Rabi}$ are either on resonance with the nearest-neighbor cavity modes at $\pm c/4\pi n_g L$ or at a higher frequency offset as in the case indicated by the blue curve in Fig. 2.

The RNGH instability gain coefficient corresponding to this refinement of the second threshold is plotted in Fig 3 as a function of the pump rate (solid blue curve). It reveals an abrupt switching-on behavior for the RNGH instability, the feature which is not seen in long-cavity QCLs because of the small frequency separation between the cavity modes (Fig. 3 red curve). The numerical results in Fig 5 are in reasonable agreement with the second threshold value of $p_{th2}$=2.35 in Fig 3 (solid blue curve) obtained using our modified second threshold condition.

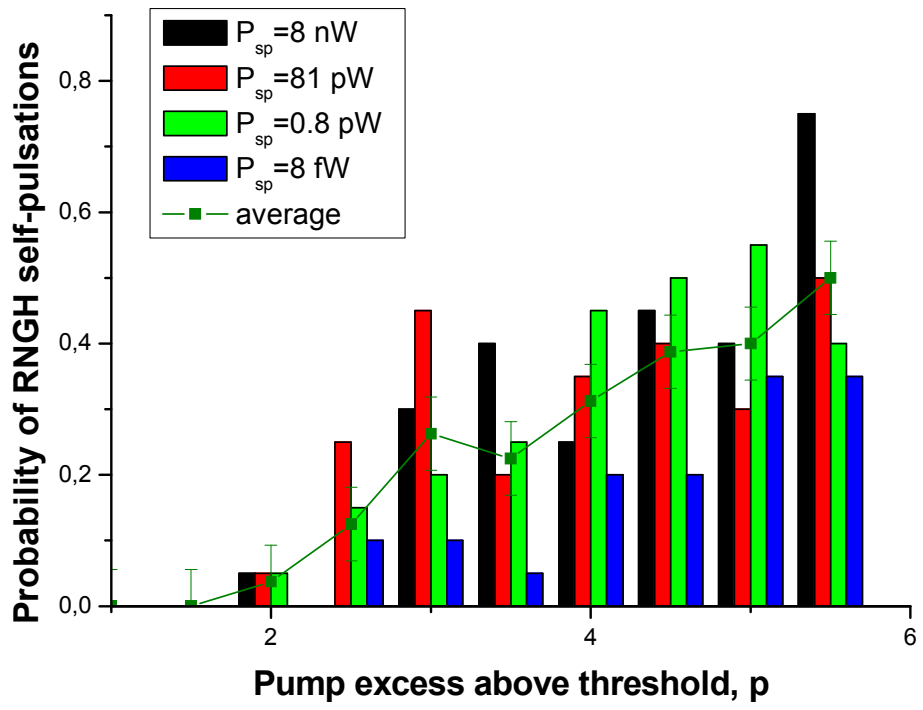

FIG. 5. Probability of occurrence for RNGH self-pulsations is plotted as a function of the pump excess above lasing threshold $p$ for different levels of spontaneous polarization noise (bars) as well as for the average over all realizations (curve). QCL cavity length is 100 µm. Other parameters are given in Table 1. See further details in the text.

The proposed refinement for the second threshold has a clear physical meaning. Recall that RNGH instability in a CW single-mode laser arises due to Rabi splitting of the lasing transition induced by the lasing mode. As a result of such spectral broadening and reshaping of the gain curve, the laser can provide sufficient optical gain for other longitudinal modes [26]. Most importantly, the medium polarization does not follow adiabatically the optical field in the cavity. Therefore in our refinement of the second threshold, we have applied the pulse area theorem [9,27], assuming that the initial perturbation to the lasing mode has a pulse shape. In the most general case, the characteristic time of such perturbation pulse is roughly a half of the cavity roundtrip time $\tau_p \sim L n_g/c$. From the pulse area theorem, we have noticed that a perturbation in the form of an optical pulse is unstable and the pulse area grows so as the fast (non-adiabatic) medium polarization dynamics becomes possible if, initially, $\Omega_{\mathrm{Rabi}} \cdot \tau_p > \pi$. Because the maximum gain for RNGH instability is located at the offset frequency $\Omega_{\max} \approx \pm\Omega_{Rabi}$, we conclude that the multimode RNGH instability occurs only when

$$\left|\Omega_{\max}\right| / 2\pi \geq c / 2Ln_g c \ . \tag{17}$$

The blue curve in Fig.2 satisfies this condition while the red one does not. This condition has minor impact on QCLs with long cavities. However it raises the second threshold considerably in short-cavity devices because of the large frequency separation between the cavity modes. In Ref.[25] we have obtained an analytic expression for the pump rate at second threshold.

### B. Experimental observations of RNGH instability in QCLs

In Section II.A we have refined the necessary conditions for occurrence of multimode RNGH instability. The numerical simulations reveal regimes with either regular or chaotic self-pulsations (sections III.C and III.D below). Therefore we should clarify which features provide an evidence for RNGH instability.

The experimental data available in the literature attest that RNGH instability does not always lead to observation of a clear Rabi splitting in the optical mode spectrum of a QCL. In some experimental realizations, RNGH instability causes just a broadening of the lasing spectrum to the offset frequencies of the order of the Rabi oscillation frequency. This is not surprising because the notion of the Rabi splitting in the gain spectrum of a laser is introduced considering small-amplitude perturbations to the initial optical field in the cavity. This picture is very different from the situation in the laser undergoing large-amplitude self-pulsations.

The appearance of Rabi splitting in the optical spectrum can be tailored by changing the sample temperature [6,16,18]. For example, in [16] at low temperatures (80-150 K) the measured lasing spectra in a buried heterostructure QCL sample are just simply broadened, but with increasing temperature, the Rabi splitting between mode clusters becomes more apparent. At room temperature conditions, two distinct mode clusters emerge in the optical spectrum. The nature of the temperature effect has not been entirely understood. The temperature-dependent saturable absorption is suggested as one of the possible mechanisms. However in the present paper we argue that low-threshold RNGH instability occurs in QCLs even without saturable absorption, due to a combined effect of coherence grating and carrier population grating.

The temperature dependent carrier diffusion leading to the contrast change of SHB grating was proposed as another possible temperature mechanism. However, for all considered temperatures, the emission spectra have exactly the same width, of ~40 $cm^{-1}$ for the pump rate of $p$~1.7-1.8 times above threshold and of ~60 $cm^{-1}$ for $p$~3.5-3.6. Both values are comparable to the Rabi oscillation frequency $\Omega_{Rabi}$. The dynamic behavior responsible for such spectral broadening is governed by very fast processes. It cannot be attributed solely to the SHB effect and mode coupling via induced grating of population inversion, whose behavior is defined by slow characteristic time $T_1$. Therefore we conclude that the large spectral broadening on the order of $\Omega_{Rabi}$ shall be attributed to multimode RNGH instability, while the optical spectrum is not necessarily split in two distinct mode clusters separated by $2\Omega_{Rabi}$. In addition, we notice that an inclusion of the coherence grating, whose relaxation time is on the scale of $T_2$ and subject to the effect of diffusion, may then provide a required mechanism to explain the thermal sensitivity of multimode RNGH instability.

Thus in [18], QCLs show a different spectral behavior with the temperature. One sample reveals two distinct lasing mode clusters with the separation of ~40 $cm^{-1}$ at liquid He temperature (6 K). However at an increased temperature of 77 K, no clear Rabi splitting is observed for all reported samples. At the same time, it was noticed that the spectral broadening and the appearance of the two distinct mode clusters is a function of the sample length. Unfortunately no detailed studies about the effect of the sample length were reported.

In the cited works, the complex structure of the lasing transition in QCL is usually not taken into account. The optical gain spectrum is usually assumed to be broad and homogeneous. At the same time inhomogeneous features in the gain curve structure or dispersive characteristics of the cavity (e.g due to backscattering from microcracks [28]) may significantly reshape the spectrum of RNGH instabilities and impact the second threshold.

Finally, even if the two distinct mode clusters are observed in the lasing spectrum, their frequency spitting is not necessary equal to $2\Omega_{Rabi}$. The pump rate dependence of the Rabi frequency [see Eq.(17) below] assumes that this splitting is proportional to the square root of the output power. In [6,16], this behavior is confirmed for the range of QCL output power up to 36 mW and the frequency splitting up to 1 THz is observed in 3 μm-wide buried heterostructure lasers. However for 10-15 μm wide ridge waveguide lasers from Refs. [16] and [18], the spectral splitting is clamped at about 0.8 THz frequency for the output power of ~25 mW and higher. Thereby we see that at high pump rates, the overall spectral broadening due to RNGH instability can be smaller than $2\Omega_{Rabi}$.

### C. RNGH instability in a QCL with long cavity

We perform numerical simulations with the TW rate equation model [22], utilizing a QCL with the cavity length of 4 mm as a model system for long-cavity devices. The linear stability analysis indicates the second threshold of $p_{th2}$=1.05 (Fig.3, red curve). For the pump rate $p$=1.2, which is slightly above the instability threshold, we do observe a quasi-periodic chaotic behavior in the output power waveform [Fig. 6(a)]. Here the cavity roundtrip time is 88 ps, much larger than the carrier lifetime $T_1$ (Table 1). In Fig. 6(a) and (b), the time domain extends over 7 cavity roundtrips. The laser is initially not pumped and the pump current is switched on at a time t=0 s. The onset of the lasing emission is seen in less than 3 roundtrips in the cavity (at t~200 ps). After ~6 roundtrips, the system reaches a steady regime of quasi-periodic (chaotic) self-pulsations. The inset in Fig. 6(a) shows a zoom at one example of such self-pulsations after 6 roundtrips in the cavity. Figures 6(c) and 6(d) show another realization obtained with the same model parameters but with the simulation domain extended to 100 cavity roundtrips. Comparing the self-pulsations in Fig.6(d) at the cavity roundtrips from 90 to 100 with those in Fig.6(c), one can see that the QCL reaches the steady regime of chaotic self-pulsations just after 6 cavity roundtrips (~530 ps). This time appears too short for establishing fixed phase relationships between many individual modes as that would be in the case of active mode-locking operation [29,30]. The last one does require thousands of cavity roundtrips. However the model laser sample we consider here does not have a built in electroabsorber section in order to produce active mode-locking operation as in [29,30], while the waveforms in Fig.6 clearly attest that this is not the mode-locking regime. Therefore the considerations about thousands roundtrips required to reach the steady operation do not apply for our single-section Fabry-Perot cavity laser.

In the original work of Risken and Nummedal [7], the steady regime of self-pulsations in a *unidirectional ring laser* is reached in less than 160 cavity roundtrips. The carrier population grating and coherence grating in our *Fabry-Perot laser* not only drastically reduce the second threshold but also reduce the time needed for settlement of the self-pulsations.

It is didactic to compare the build-up time of the lasing emission in Fig.6 with an estimate based on the well-known text book expression that reads $\tau_c \cdot \ln(P_{LAS}/P_{SE})/(p-1)$ (see e.g. [31]). For the case considered in Fig.6, the photon lifetime in the cavity is $\tau_c$=4 ps while the power ratio in the steady lasing regime and in the initial regime of spontaneous emission (at $t$=0) is of $P_{LAS}/P_{SE}=10^6$ [see Fig.6(e)]. For the pump excess above threshold of p=1.2, the estimated build-up time is of 280 ps, which corresponds to 3 roundtrips in the cavity. This estimate agrees very well with the rise time of the waveforms in Fig.6.

Once the lasing is reached, the roundtrip gain coefficient for the multimode instability at p=1.2 becomes of 0.38 per cavity roundtrip (see also [25]). That is, the amplitude of an unstable non lasing mode increases by a factor of $\exp\{0.38N\}$ after $N$ roundtrips. This process has the characteristic time scale of 1/0.38= 2.6 cavity roundtrip. Therefore reaching the steady regime of multimode self-pulsations just in additional 2-3 roundtrips after the onset of lasing emission (by t~500 ps in Fig.6) appears very reasonable.

In Fig. 6(b) we plot the system attractor, which traces the evolution of the medium polarization *P* versus carrier density *N*. Both *P* and *N* values are taken in the vicinity of the laser output facet and are normalized at transparency carrier density. The P-N attractor has a characteristic butterfly shape typical for a chaotic behavior in a Lorenz-type system [32]. We observe similar chaotic attractors with increasing pump power.

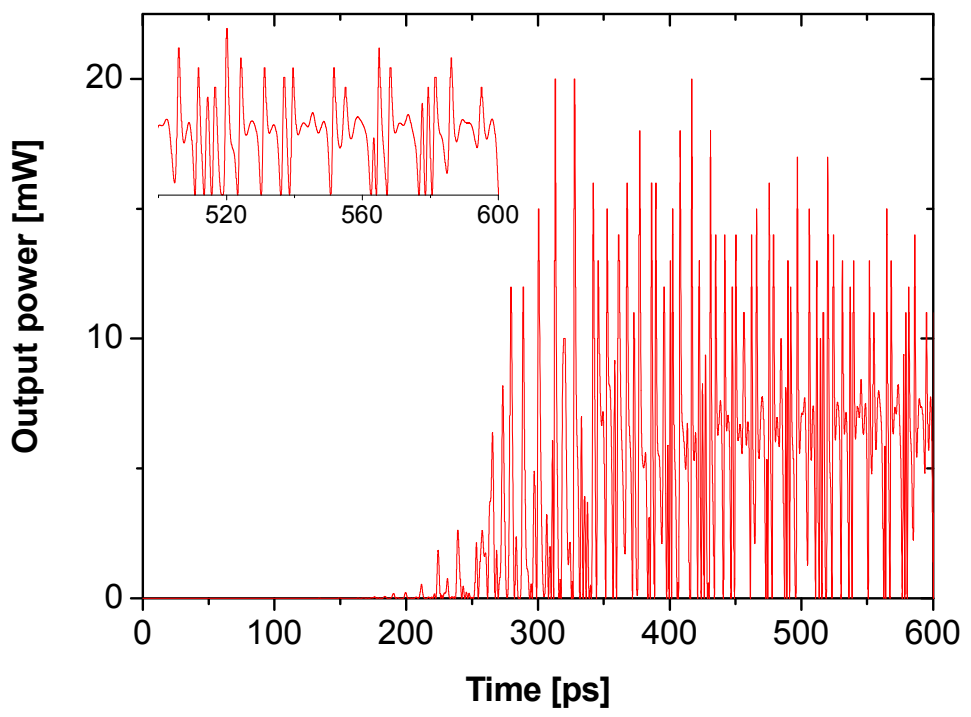

(a)

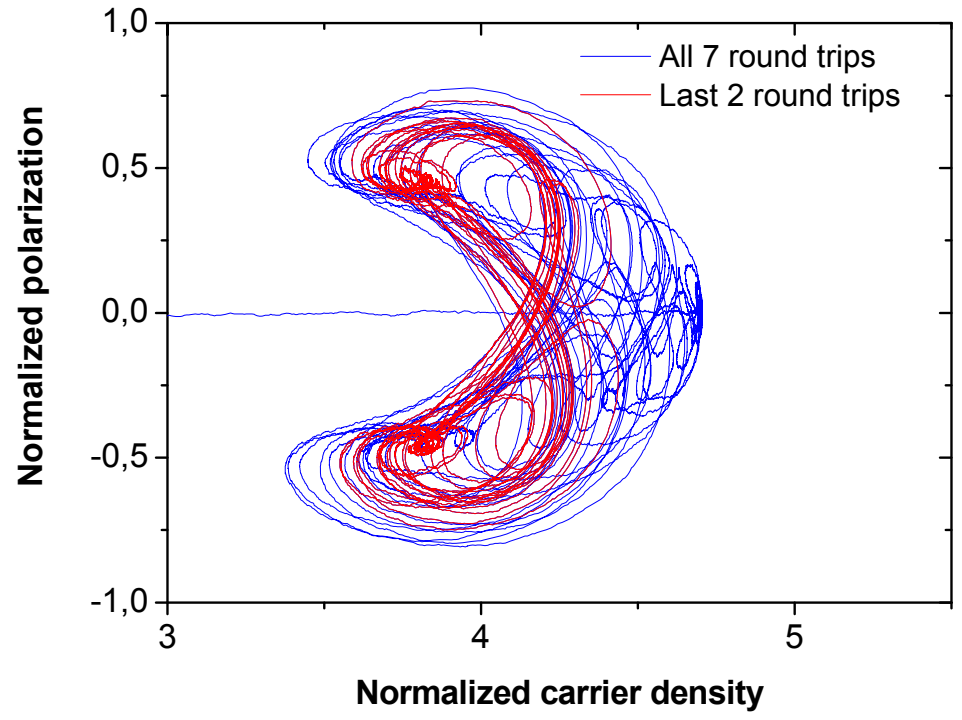

(b)

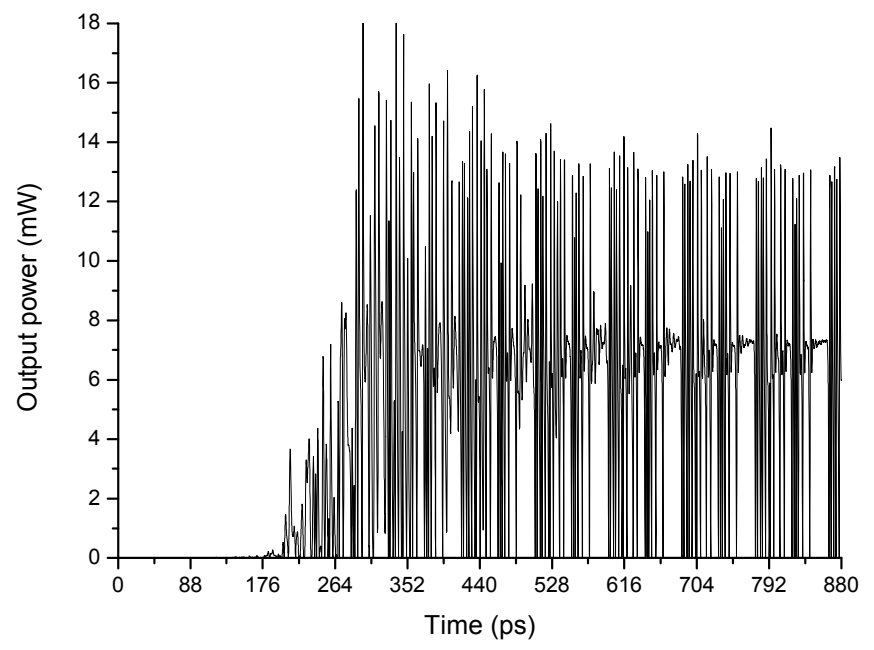

(c)

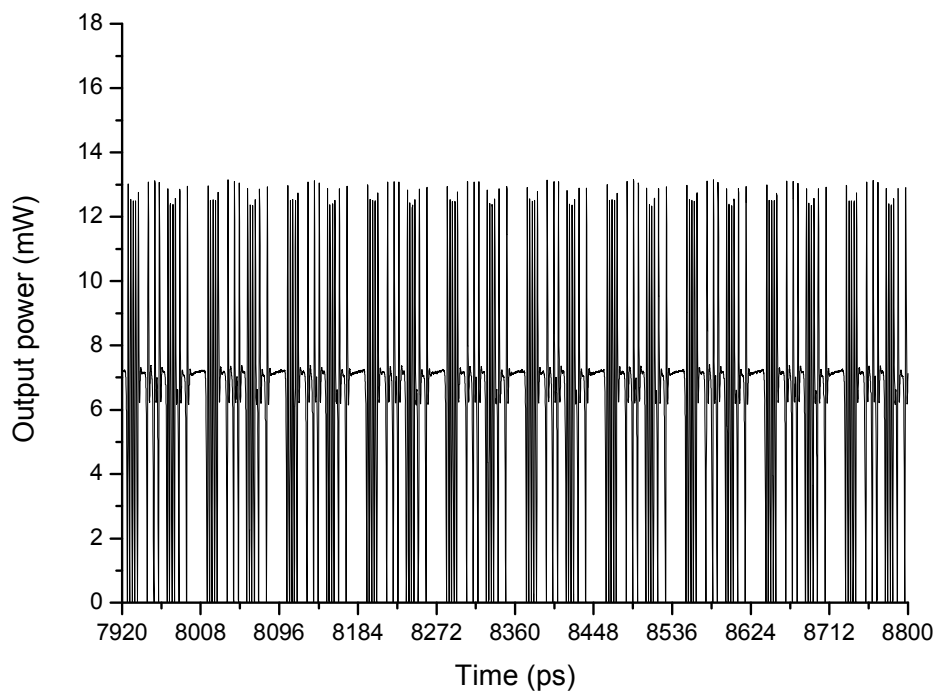

(d)

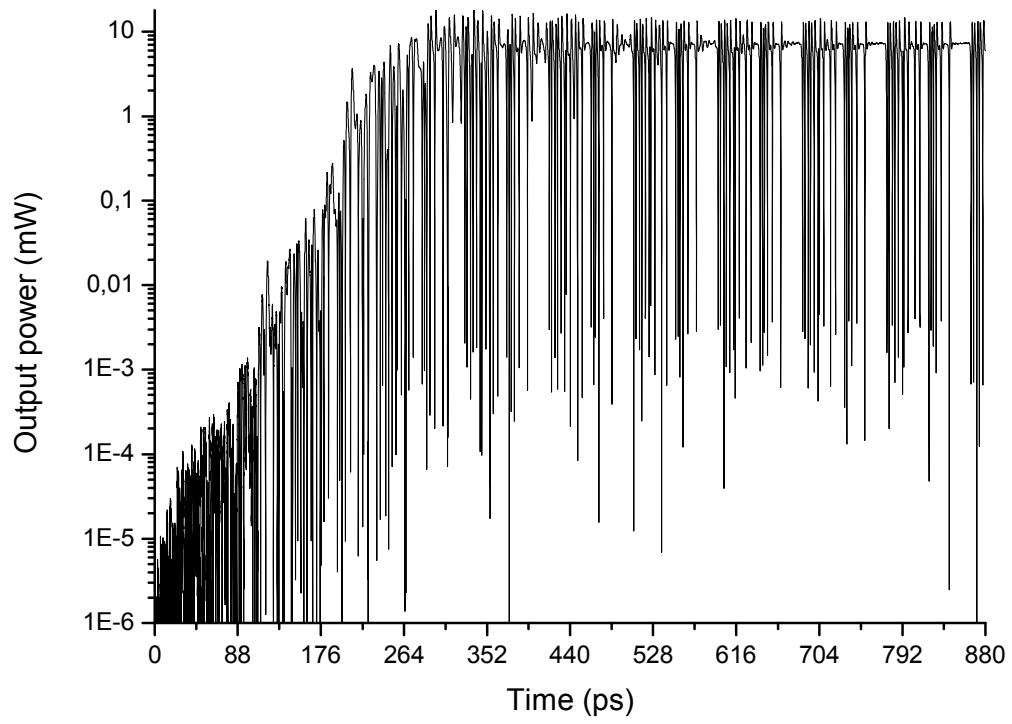

(e)

FIG. 6. Results of numerical simulations with TW model for QCL with the cavity length of 4 mm (cavity roundtrip time is 88 ps): (a) Plot of the output power as a function of time. The inset spans over last two roundtrips in the cavity. (b) Chaotic P-N attractor. The behavior during all seven cavity roundtrips is depicted with the blue curve. The red curve signifies the last two roundtrips, when the system reaches the steady regime of quasi-periodic (chaotic) self-pulsations. Panels (c) and (d) show another realization obtained with the same model parameters but with the simulation domain extended over 100 cavity roundtrips, depicting, respectively, the first ten and the last ten roundtrips in the cavity. The steady regime of quasi-periodic self-pulsations as seen after 90 roundtrips in (d) is actually reached after ~6 cavity roundtrips (~530 ps) as seen in (c). The panel (e) shows the details of the fast build-up of the lasing regime on a logarithmic scale. The QCL is pumped at p=1.2 times above the lasing threshold. Other parameters are shown in Table I.

With increasing pump rate, the optical field waveform develops into a quasi-periodic square wave. An example of the temporal behavior of the optical field amplitude and medium polarization for a wave traveling in the positive z axis direction as well as the carrier density evolution is depicted in Fig. 7. These waveforms are obtained for the pump rate of $p$=2.2. The optical field behavior [Fig. 7(a)] appears to be quite similar to the one predicted in Ref [23] for a long-cavity unidirectional ring laser at high pump rates. However in our case, the average field amplitude is zero due to quenching of the initially CW lasing mode and emergence of the two symmetric sidebands in the optical spectrum.

Like the optical field, the medium polarization shows the square-wave behavior. However its waveform pattern is not identical to that one of the optical field [Fig. 7(b)]. That is not surprising if one takes into account the scattering of counter-propagating waves on the induced gratings of carrier coherence and population. Most importantly, the medium polarization does not follow the optical field adiabatically. Instead, the medium polarization itself defines the optical field dynamics. One can see in Figs. 7(a) and 7(b) that the pattern of the square wave almost repeats itself after each roundtrip in the cavity, indicating that a wave packet with complex envelope travels back and forth in the cavity. Its envelope just slightly changes at each roundtrip, yielding the quasi-periodic chaotic behavior.

A different pattern is seen in the waveforms of the optical power and carrier population [Figs. 7(c) and 7(d) respectively]. These variables are either quadratic with respect to the optical field or have a contribution from the product of the optical field amplitude and the medium polarization [see Eqs.(7)-(11)]. Each waveform in Figs. 7(c) and 7(d) exhibits spikes that are superimposed on a steady level. The spikes are caused by intermittent behavior of the optical field and polarization, while the steady-state component attests for the symmetry of the field and polarization square wave patterns. The period of spikes is roughly a half of that for polarization waveform, yielding the Lorenz-type attractor in Fig. 6(b). All these features seen in the time-domain waveforms are directly imprinted into the optical and radio frequency (RF) spectra. In Fig. 8 we show evolution of these spectra with the pump rate and compare these with the Rabi oscillation frequency.

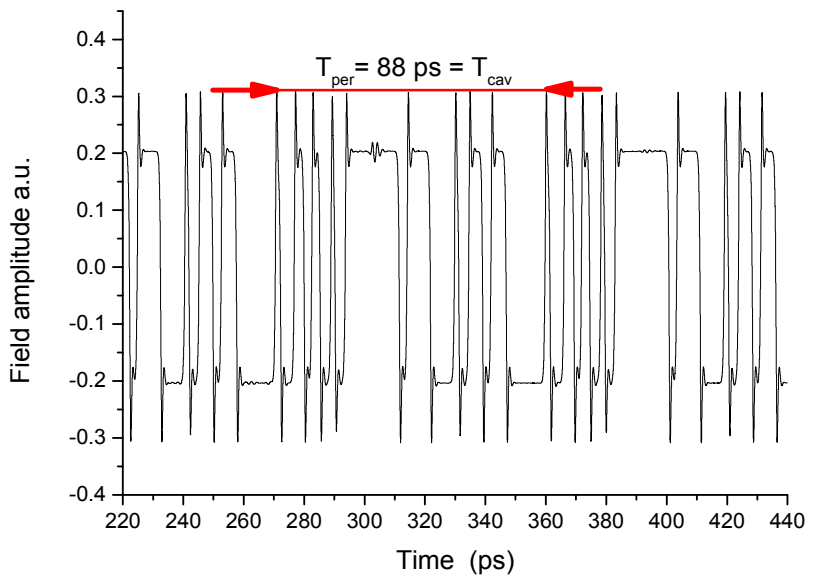

(a)

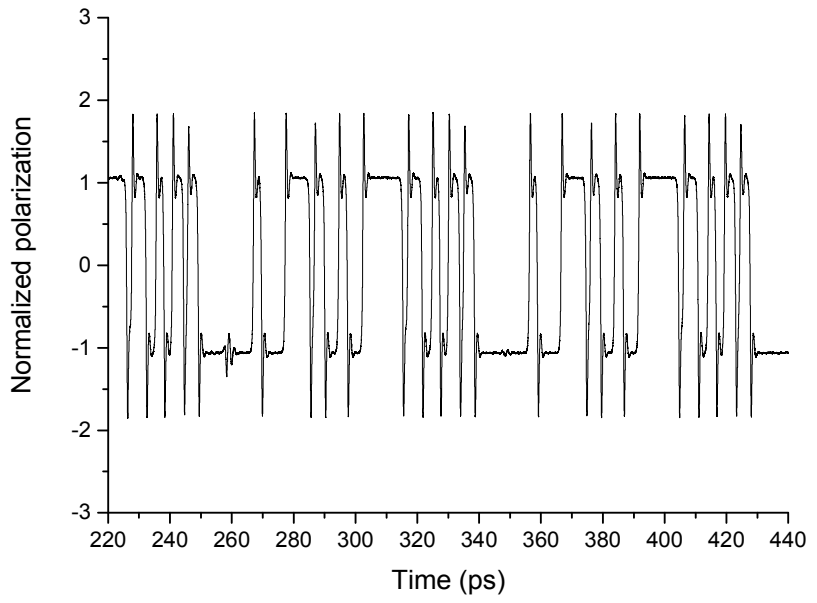

(b)

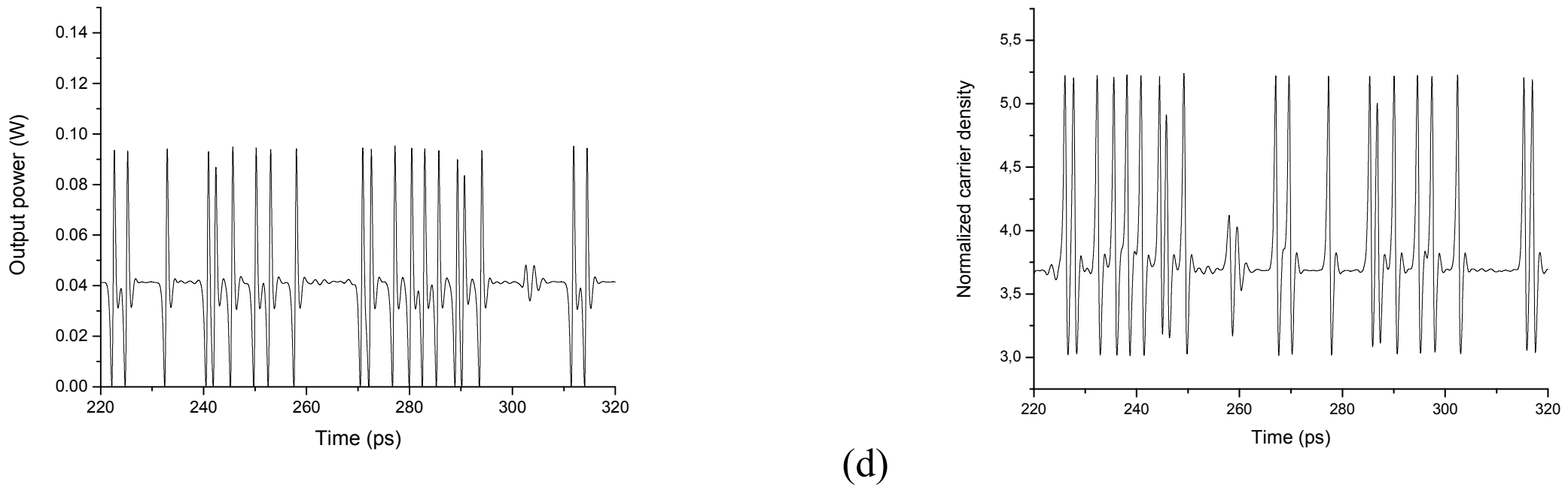

(c) (d)

FIG. 7. Results of numerical simulations with TW model for QCL with the cavity length of 4 mm: The optical field waveform for the wave propagating in the positive z axis direction (a) and normalized medium polarization associated with this wave (b), output power (c) and normalized carrier density (d). All values are taken nearby the output facet of the laser cavity. The QCL model parameters are the same as in Fig. 6 but the pump rate is of p=2.2.

Due to the standing wave pattern of the optical field in the cavity and induced carrier coherence and carrier population gratings, there are several Rabi oscillation modes in the cavity. In order to extract their frequencies, we have made an additional small-signal analysis of Eqs.(7)-(11). This time, we keep fixed the amplitude of the optical standing wave in the cavity. We find that Rabi oscillations are possible at the main frequency $\Omega_{Rabi}$ as well as at the two other frequencies $(1+2^{-1/2})^{1/2}\Omega_{Rabi}$ and $(1-2^{-1/2})^{1/2}\Omega_{Rabi}$, where:

$$\Omega_{Rabi} = \sqrt{\frac{p-\nu_0(p)}{T_1 T_{2_eff}}} \tag{18}$$

Out of these three eigen modes, only the main one (18) is associated with the gain medium variables in the lower 4×4 matrix block of Eq. (16), the one which is responsible for RNGH instability. In Fig. 8, we also compare the extracted spectra with the evolution of the peak gain frequency $\Omega_{\max}$ for multimode RNGH instability (see Fig. 2) as well as with the possible frequency of multimode instability due SHB effect, which is predicted in [16] to be well below $\Omega_{Rabi}$ and $\Omega_{\max}$:

$$\Omega_{SHB}^{[16]} = \sqrt{\frac{1}{T_1}\sqrt{\frac{p-1}{3T_1T_2}}} \,. \tag{19}$$

All these frequencies ($\Omega_{Rabi}$, $\Omega_{\max}$, and $\Omega_{SHB}^{[16]}$) are obtained using a small-signal approach. Because the optical power is $\propto E^2$ while the carrier dynamics is governed by terms $\propto EP$, in Figs. 8(a) and 8(b) we plot these frequencies scaled by a factor of 2. Note that the peak gain frequency $\Omega_{\max}$ is quite close to the Rabi oscillation frequency (compare the red and blue curves in Fig. 8).

The evolution of the RF power spectrum in Fig. 8(a) almost follows along the curves for $2\Omega_{Rabi}$ and $2\Omega_{\max}$ frequencies, exhibiting the spectral broadening and the frequency shift of the modulation band that increases with the pump rate. Note however that it is unrealistically to expect the exact matching between the numerical simulations for large-amplitude self-pulsations and the outcomes of the small-signal analysis. At the same time the spectral broadening and the frequency shift in Fig. 8(a) are clearly much larger than for a possible multimode instability at the frequency $\Omega_{SHB}^{[16]}$ due to SHB effect [16] (green curve). Therefore we attribute the spectral behavior in Fig. 8(a) to RNGH-like instability. Since the output power waveform as in Fig. 7(c) can be regarded as a series of ultrafast spikes superimposed on a steady intensity level, the modulation band seen in the RF power spectrum at the frequency $2\Omega_{Rabi}$ is primarily due to the spiking behavior of the output power.

In Fig. 8(b) we plot the frequency-domain representation of the carrier density dynamics at different values of $p$. The spectra show broadening and modulation of the carrier density at frequencies up to twice the Rabi oscillation frequency. This behavior is similar to the evolution of RF spectra in Fig. 8(a). However the modulation band nearby $2\Omega_{Rabi}$ is less pronounced and the spectra exhibit almost all frequency components down to $\Omega$=0. This behavior can be attributed to a slow response of the carrier density and carrier population grating to rapid variations $\propto EP$ in Eqs.(10)-(11). The cutoff frequencies are of $1/2\pi T_1 = 122$ GHz and $1/2\pi T_g = 172$ GHz for the average density and population grating, respectively.

The medium polarization waveforms in the frequency domain representation and the optical spectra are displayed in Figs. 8(c) and 8(d), respectively. As expected, the two sets of spectra show very good resemblance. In both sets, the initially lasing mode at $\Omega=0$ (the carrier wave) is quenched and the spectra reveal two symmetric sidebands. Since the RF modulation spectra of the output power exhibit modulation bands at the frequency $2\Omega_{Rabi}$ [see Fig. 8(a)] one would expect to observe the modulation sidebands at frequencies $\pm\Omega_{Rabi}$ in the optical spectra as well. However these considerations do not take into account large phase variations due to intermittent behavior of the optical field, as seen in Fig. 7 (a). Obviously, the phase of the optical field does not contribute to the RF power spectrum. At the same time, the large phase hops of $\pm\pi$ impact directly the overall optical spectrum, warping its envelope as compare to the envelope of the RF power spectrum. From the waveform in Fig. 7 (a) one can see that the average repetition frequency of such phase hops is much lower than the spectral band of each spike in Fig. 7(c). As a result, the modulation sidebands in the optical spectrum are shifted to lower frequency. Therefore one of the possible origin for experimentally reported clamping of the optical spectrum bandwidth (see Sec. III.B) is that the optical field waveform develops into a quasi-periodic square wave, showing the intermittent behavior.

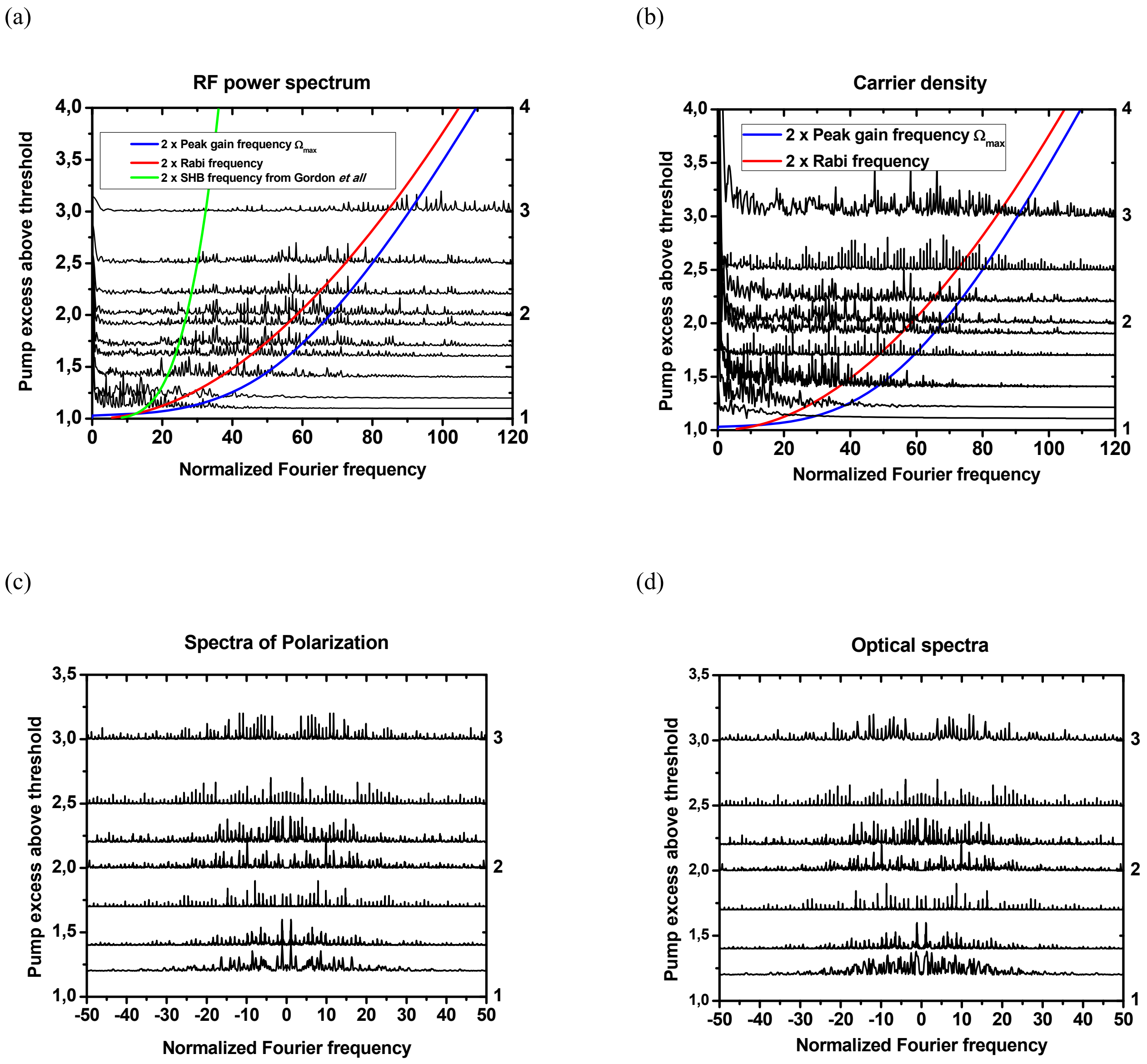

FIG. 8. Results of numerical simulations with TW model for QCL with the cavity length of 4 mm: Evolution of QCL (a) RF power spectrum, (b) carrier density, (c) polarization and (d) optical spectrum when pumped above the lasing threshold. Red and blue curves indicate expected location of modulation sidebands originating from $\Omega_{\mathrm{Rabi}}$ and $\Omega_{\mathrm{max}}$. The green curve shows that one for $\Omega_{SHB}^{[16]}$ . The frequency is normalized to longitudinal mode spacing of 11.4 GHz. The model parameters are the same as in Figs. 6 and 7.

### D. RNGH instability in a QCL with short cavity

We now move on to the case of short-cavity QCLs and consider an example of QCL with the cavity length L=100 μm, for which the linear stability analysis predicts the second threshold of $p_{th2}$= 2.35 (see Fig.3, solid blue curve). The results of TW model simulations for the pump rate of p=1.5, which is below the second threshold, are shown in Figs.9 (c) and (d). After several roundtrips in the cavity, the laser reveals an onset of damped relaxation oscillations and a transit to CW lasing regime, as can be seen from both output power waveform [Fig. 9(a)] and spiral-shape P-N attractor [Fig. 9(b)]. The model simulations are thus in agreement with the predictions of the linear stability analysis in Fig. 3.

The QCL reveals a different behavior when pumped above the second threshold. At p=2.5, the first emission burst of high peak power is followed by steady regime of regular self-pulsations [Fig. 9(c)]. The FWHM pulsewidth of the first emission burst is of 0.5ps while for regular self-pulsations it is of 0.6 ps width. Note that these self-pulsations are much faster as compare to the relaxation oscillations at below second threshold in Fig. 9(a). The system attractor plotted in the P-N plane indicates that the laser transits to regular self-pulsations just after a few roundtrips in the cavity [Fig. 9(d)] once it is at lasing. Indeed in Fig.9, the gain coefficient for multimode instability is of 0.5 per roundtrip, yielding the characteristic time scale of just two cavity roundtrips. The last portion of the time domain used in the numerical simulation is depicted in Figs. 9(e) and (f). Note that the period of regular self-pulsations is close to the cavity roundtrip time of 2.2 ps while the optical field amplitude (and the medium polarization) changes sign at each half period (This behavior attests that the considered regime is not a usual mode-locking). This can be seen from the 8-like shape of P-N attractor in Fig. 9(f).

Figure 10 shows the optical and RF power spectra when the pump rate *p* is in the range of 2.2 to 3.9 times above the lasing threshold. In case of short-cavity QCL operating at just above the second threshold, only two main frequency components are excited in the optical spectrum [Fig. 10(a)]. With increasing pump rate, the spectrum remains symmetric with respect to the initially lasing mode and the two main spectral components are located at the two nearest-neighbor modes. Respectively, the RF power spectrum shows the main harmonic at about twice of the cold cavity mode spacing [Fig. 10(b)]. As discussed in Sec.III A, the multimode RNGH instability occurs when the frequency $\Omega_{max}$ of the maximum gain for instability is on resonance with the first adjacent cavity mode [Fig. 10(a), blue curve]. However, with increasing pump rate, the frequency of self-pulsations does not follow the increasing frequency $\Omega_{max}$ or Rabi oscillation frequency (red curve). We attribute this behavior to the cavity roundtrip self-repetition condition (see Sec. II.C). Instead of the continuous frequency rise, higher order modes start to appear in the optical spectrum. Note that the frequency of self-pulsations is slightly lower than the cold cavity mode spacing. The group velocity reduction can be attributed to propagation phenomenon of a high-energy pulse in resonant medium, the process which can be regarded as continual absorption of energy from the pulse leading edge and re-emission of energy into the pulse trailing edge [27].

A quite similar behavior is reported in [23] for a short-cavity unidirectional ring laser. However in case of the ring laser, the optical field does not change the sign at each half-period and, as a consequence, the main lasing mode at $\Omega=0$ is not quenched. One can trace a few other similarities between multimode instability in QCLs discussed here and the RNGH instability in unidirectional ring lasers from Ref. [23]. In particular, one very important question has been challenged but has not been answered in [23]. More specifically, the mechanism, which is responsible for appearance of regular self-pulsations in case of short-cavity laser instead of chaotic pulsations in the long-cavity case, have not been elucidated. This qualitative change in dynamic behavior becomes even more wondering if one consider the affinity between the 8-shape attractor in case of regular self-pulsations in Fig. 9(f) and the butterfly-shape attractor in case of chaotic self-pulsations in Fig. 6(b).

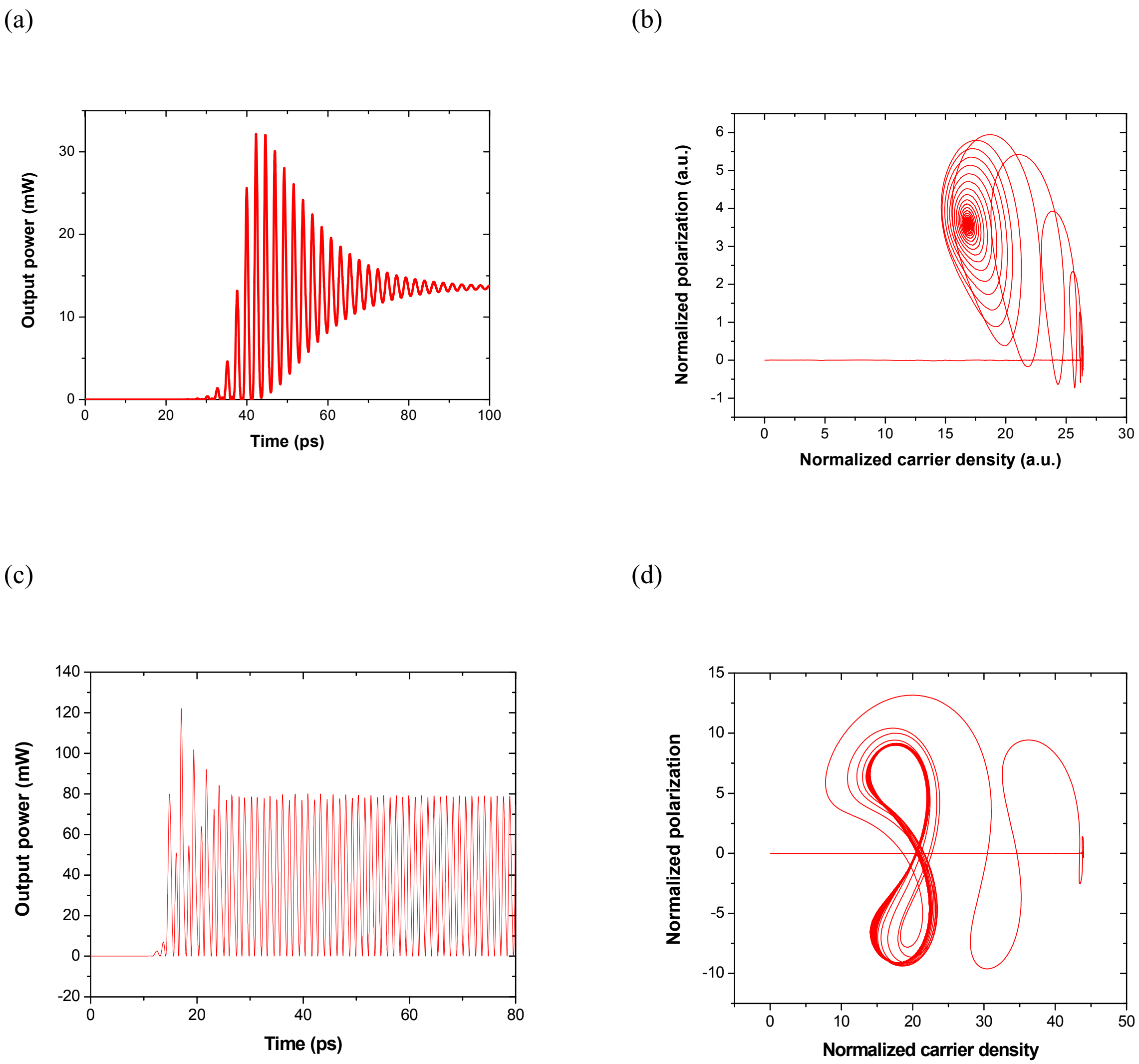

(e) (f)

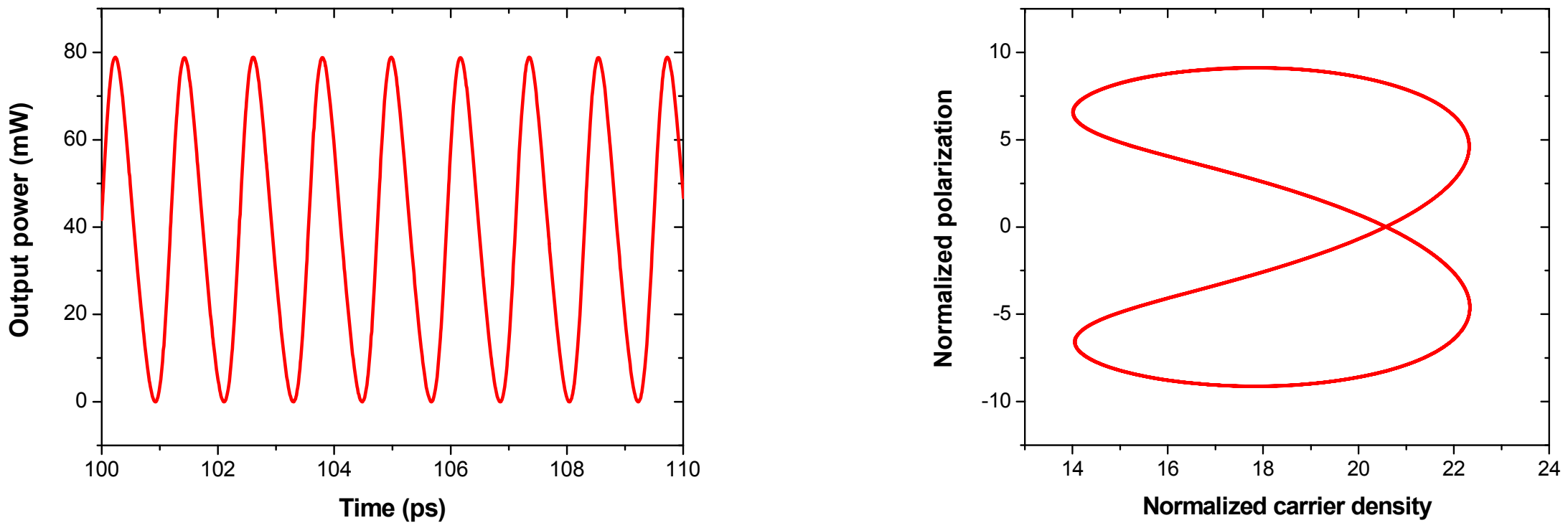

FIG. 9. Results of numerical simulations with TW model for QCL with the cavity length of 100 μm: Waveform (a) and P-N attractor (b) in case of p=1.5 (below 2nd threshold), the system is stable. Waveform (c) and P-N attractor (d) for p=2.5, zoom- in at the end of simulation domain which shows regular self-pulsations with sub-picosecond pulse width (e) and P-N attractor (f).

Below we propose one possible explanation on the effect of the cavity length. To get an insight into its cause we study the P-N attractors in further details. The P and N variables used here are introduced following the approach of Ref. [22]. In particular, the variable P measures the order parameter in the system (the carrier density in the coherent state) and has the same units as the carrier density N. In Figs. 6 and 9, both variables P and N are normalized on the transparency carrier density.

In Fig. 9 the order parameter P is high. It reaches about a half of the value for N that precedes the emission pulse when P=0. This feature is observed as in the first emission burst [Fig. 9(d)] as well as in the regime of regular RNGH self-pulsations [Fig. 9(f)]. This behavior indicates an "off-diagonal long range order" in the system, as predicted in the pioneering paper by Graham and Haken [8]. Note that a similar behavior is also predicted for Dicke superradiance (SR) [33,22]. The similarity is even far more striking. Thus as shown in [8], the master equation for RNGH instability behavior can be put in the form that bears a closed formal analogy to the description of condensation phenomena, such as superconductivity. But in [22] (and also in [34]), the SR emission is also shown to be governed by the master equation in the form of Ginzburg-Landau equation. Continuing these parallels we note that in the pioneering work of Risken and Nummedal [7], the approximate analytic solution for the regime of RNGH self-pulsations was obtained in the form of a hyperbolic secant pulse. This was achieved by considering the case, in which the optical filed is entirely defined by the medium polarization, so as $E \propto P$. The same relationship between $E$ and $P$ as well as the same hyperbolic secant pulse shape applies to the case of Dicke superradiance [22,35]. There is however one important difference between the two regimes. The RNGH instability is usually analyzed in a CW operating laser when the active medium is under continuous wave pumping. The SR pulse emission occurs from a strongly pumped active medium, when the initial optical field is close to zero. In practice this is achieved with pumping by short and intense optical or electrical pulses [33,35,36,37,38,39]. Taking all these considerations into account, we attribute the first emission burst in Figs. 9(c) and (d) to SR.

It was shown that SR in short samples is different from the cooperative emission in long samples [22,34,40]. The borderline between the two cases is set by the coherence length of the SR pulse. In case of long samples, the situation is such that sample domains of the size of the coherence length emit independently of each other. The output pulse is the result of incoherent superposition of SR emission from different sample domains. As a result, the overall pulse width broadens and its amplitude decreases.

In a similar way, we may conjecture that the regular RNGH self-pulsations occur when the sample is shorter than the coherence length. Otherwise, in the case of a long sample, different sample domains are not mutually coherent with respect to instable cavity modes and compete with each other at the initial stage. As a result, an erratic pattern of the field is established in the cavity at a later stage. As seen in Fig. 7, the optical field pattern almost repeats itself on subsequent roundtrips in the cavity, however the coherence length remains smaller than the cavity length. It is roughly the average time between the phase hops in Fig. 7(b).

A more detailed investigation about the cavity length effect on the coherence length of RNGH self-pulsations is left for another study.

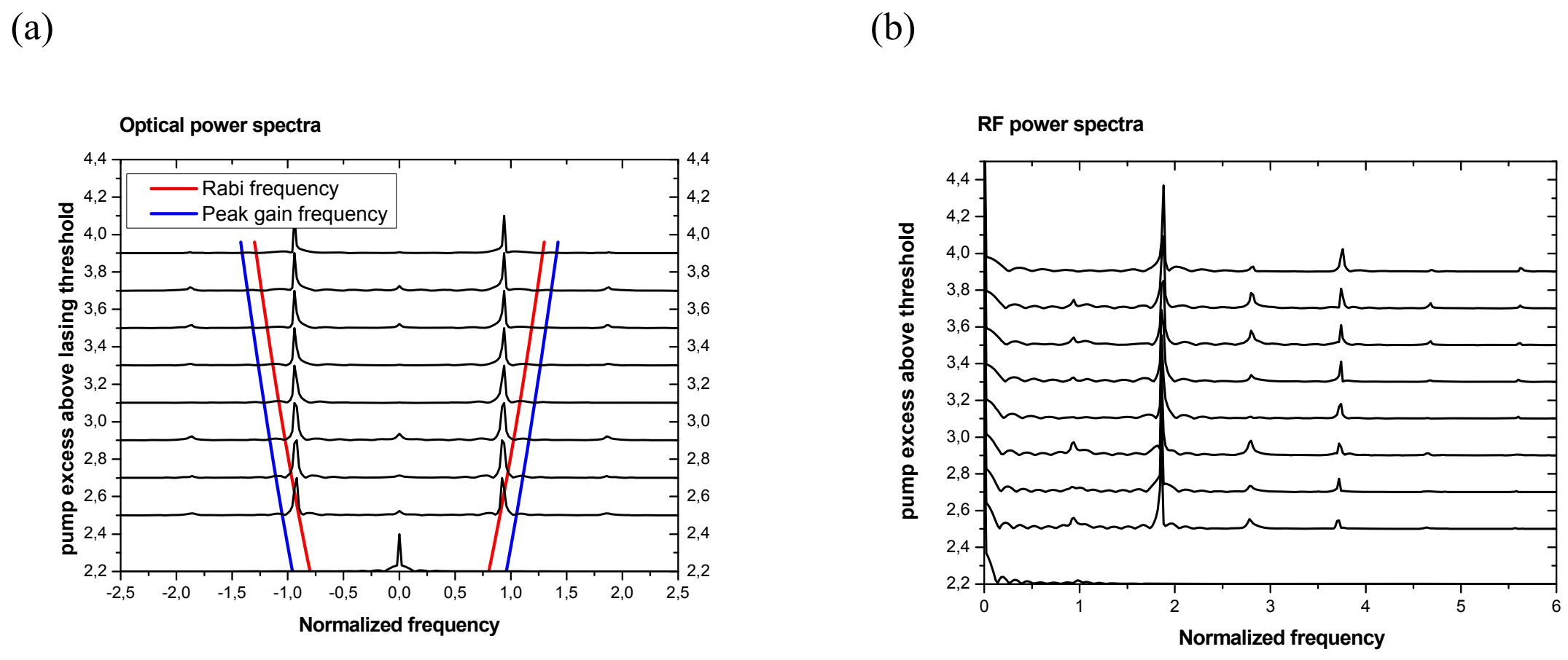

FIG. 10. Results of numerical simulations with TW model for QCL with the cavity length of 100 μm: Optical (a) and RF power (b) spectra.

## E. The role of carrier diffusion

We have shown in Sec.II.C that the carrier coherence grating and carrier population grating are both responsible for lowering the multimode instability threshold. The diffusion coefficient D, which affects their effective relaxation times $T_g$, $T_{2_g}$ and $T_{2_eff}$, is sensitive to the temperature. Unfortunately the literature data about the temperature effect on RNGH self-pulsations are quite controversial (see Sec. III.B), which is likely because the temperature also affects many other QCL parameters. On the other hand, the relaxation rates enhancement due to the carrier diffusion is proportional to the square of the photon wavenumber $k$ [see the discussion to Eqs.(7)-(11)]. Therefore we elucidate the role of the carrier diffusion by considering various emission wavelengths.

First we consider the case of QCLs emitting in the MIR spectral range and discuss the effect of electron diffusion in the plane of active QWs in these devices. In Fig. 11 we compare the increments for multimode RNGH instability in QCLs samples emitting at 4 µm (green curve) and 10 µm (red curve) wavelengths. Without carrier diffusion, the second threshold is unrealistically low, at below p=1.005 in long-cavity samples (black dash-dotted curve). The relaxation due to diffusion rises the second threshold. As expected, this effect is more pronounced for shorter wavelength QCLs, where relaxation is faster (compare red and green curves). Thus for devices with the cavity length of 4 mm, the second threshold $p_{th2}$ is of 1.04 at the wavelength of 10 µm, while $p_{th2}$=1.32 in QCL emitting at 4 µm. The effect of the diffusion on the second threshold is seen as in the long cavity devices (solid curves) as in the short cavity QCLs (vertical dashed lines). In all considered cases, the carrier diffusion does not rise the second threshold to prohibitively high levels.

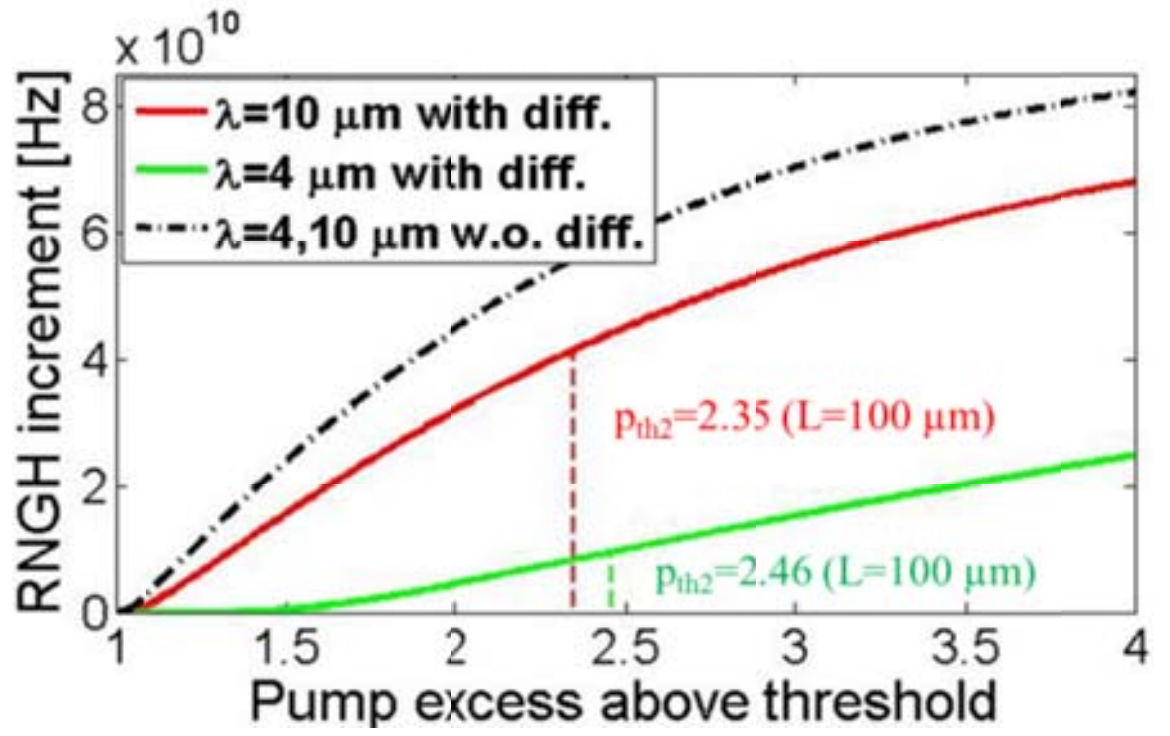

FIG. 11. Maximum RNGH instability increment vs pump excess above threshold for QCLs without relaxation due to carrier diffusion (black dash-dotted curve) as well as with the effect of the carrier diffusion when the emission wavelength is 4 µm (green curve) or 10 µm (red curve). The solid curves are obtained for long-cavity devices (L=4 mm), while the vertical dashed lines denote the second threshold for short-cavity devices (L=100 µm). Other parameters are indicated in Table I.

The implication of the carrier diffusion effect is totally different in the case QW laser diodes operating in the VIS or NIR spectral range. Even though the ambipolar diffusion in the QWs of these devices is much weaker than the diffusion of electrons in QCLs (see Table I), the relaxation due to diffusion is more perceptible because of the shorter wavelength and smaller spatial period of the induced gratings. It is commonly acknowledged that QW laser diodes with monolithic FP cavities do not show RNGH instabilities in the range of pump currents that can be reached in practice.

In Fig. 12, we study the increment for multimode RNGH instability in a FP-cavity GaN semiconductor laser diode operating at 420 nm and in a GaAs LD at 850 nm wavelength. In all cases examined in Fig. 12, the two LDs show quite similar behavior. In Figs. 12 (a) and (b) we consider, respectively, the eigen solutions of the 4×4 and 5×5 matrix blocks in Eq.(16) and plot the largest real part of the Lyapunov exponents. The first matrix is related to the instability caused by the carrier population and carrier coherence gratings, which is the focus of this paper. The second matrix would reproduce the results of the original RNGH theory if both induced gratings are removed ($T_g, T_{2_g} \to 0$) while the carrier relaxation and carrier dephasing processes preserve their initial times scale ($T_1$ and $T_{2_eff}$). Recall that the RNGH theory [7,8] was originally established for *a unidirectional ring laser*, where the gratings cannot be formed. It predicts the second threshold $p_{th2}$ at more than 9 times above the lasing threshold. The second threshold of the original RNGH theory would be recovered by our model if, in particular, $T_{2_g}/T_{2_eff} \to 0$ (see section III.F below). However, neither with carrier diffusion ($T_{2_g}/T_{2_eff} \sim 0.5$ see Table 1) nor without it ($T_{2_g} = T_{2_eff}$), the VIS-NIR LDs cannot reach this condition. As a result, the second threshold emerging from the 5×5 matrix block in Eq.(16) is very high in the VIS-NIR LDs (Fig. 12(b)) due to the presence of induced gratings.

The carrier coherence and carrier population gratings have a different implication on the instability emerging from the 4×4 matrix block [Fig.12(a)]. Thus in a *FP cavity lasers* as we consider here the induced coherence and population gratings lower significantly the second threshold. However without the ambipolar diffusion of carriers, the model predictions for the second threshold in VIS-NIR LDs would be unrealistically low, of $p_{th2} \approx 1$ (dash-dotted curves in Fig. 12). The relaxation due to the carrier diffusion renders our model to be more realistic (solid curves in Fig. 12(a)) because the characteristic time constants $T_1$=1 ns and $T_2$=0.1 ps are now reduced to much shorter effective relaxation times of $T_g$=0.1-0.2 ps for the carrier population grating, $T_{2_g}$~0.05ps for the coherence grating and $T_{2_eff}$ ~0.09ps for the effective dephasing rate. This lifetime reduction strongly suppresses all effects induced by the standing wave pattern of the optical field in the cavity. As a consequence, the second threshold values predicted from the linear stability analysis are of several hundred times above the lasing threshold, as indicated in Fig. 12(a). Recall that this analysis is based on the truncated set of coupled-mode equations (7)-(11) and the model predictions for the second threshold cannot be quantitatively accurate at such a high pump rate. Nevertheless these predictions are in a reasonable agreement with the outcomes of our numerical simulations based on the TW rate equation model, which we discuss next.

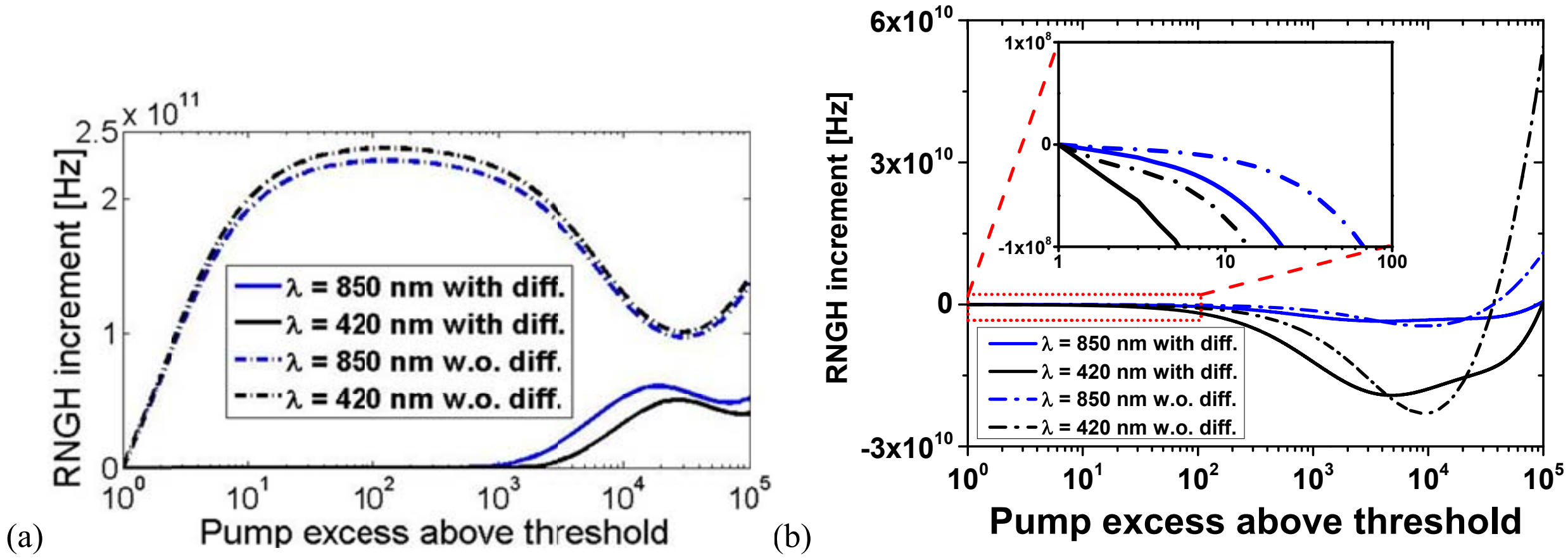

FIG. 12. (a) Maximum RNGH instability increment from the 4×4 matrix block in Eq (16) and (b) largest real part of the Lyapunov exponent from the 5×5 matrix block vs pump excess above the lasing threshold for GaN LDs (black curves) and GaAs LDs (blue curves), calculated with and without carrier diffusion (solid curves and dash-doted curves respectively) calculated from (a) 4x4 and (b) 5x5 matrix block. Inset in (b) shows zoom at p from 1 to 100. The cavity length is L=4 mm, other parameters are shown in Table I.

In Fig. 13 we show the output waveform and P-N attractor simulated numerically with TW model for a 100 μm long single-section GaN LD pumped above the second threshold, at p=400. At such high pump rate, LD exhibits behavior that is similar to QCLs biased at above the RNGH instability threshold. The output waveform at each cavity facet reveals the SR emission burst followed by regular RNGH self-pulsations [compare to Fig. 9(c)]. Note that during the SR emission burst, the order parameter is large and approaches a half of the initial carrier density stored in the system. At the pump rates below second threshold, LD shows an ordinary behavior of a Class-B laser, which is characterized by excitation of damped relaxation oscillations followed by a transition to CW lasing regime [as in Fig. 9(a)].

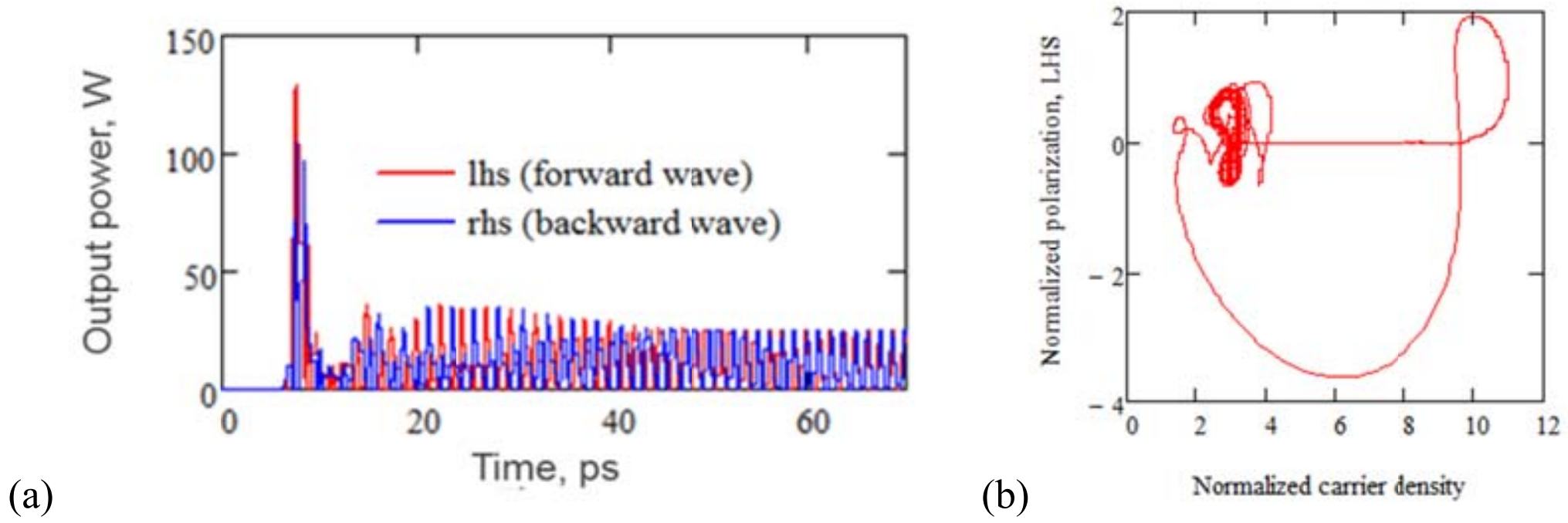

FIG. 13. Results of numerical simulations with TW model for GaN LD with the cavity length of 100 μm: Waveform (a) and P-N attractor (b). The red (blue) trace corresponds to the wave propagating in the forward (backward) direction.

We thus conclude that VIS-NIR range QW LDs have very high second threshold because of rapid relaxation of the carrier coherence and population gratings. Whereas the RNGH instability threshold in single-section QW LDs is not reachable under realistic experimental settings, there are several considerations that indirectly support our conclusion.

Reaching the SR emission in short-length mesa-etched structures comprising GaAs QWs was attempted in [41] under short-pulse optical pumping. However no evidence of SR emission was observed. We may attribute this to the fact that threshold condition for RNGH instability has not been reached in these experiments as the pump rate was too low.

According to Fig. 12, lowering of RNGH instability threshold (and reaching SR emission) in single-section LD can be achieved by reducing the contribution of diffusion to the relaxation rates of the carrier coherence and population gratings. This bring us to the insight that low-dimensional semiconductor heterostructures such as quantum dots (QDs) or quantum dashes (QDash) can be used as active gain material in order to avoid prohibitively high second threshold. Surprisingly, in Ref. [42], starting from totally different considerations, a similar conclusion was made about semiconductor heterostructures suitable for SR emission. Thus according to [42], SR emission is not possible with the active gain medium utilizing bulk semiconductor material or QW heterostructures. One needs to use QDs or to introduce an additional quantization degree in a QW by applying a strong magnetic field. The SR emission from magneto-excitons in InGaAs/GaAs QW was confirmed in Ref.[35] under short-pulse optical pumping.

There is another way to reduce the second threshold in a semiconductor laser. It consists in incorporating a saturable absorber [17] (see also Introduction and Section II.C). Technically this is achieved by implementing several separately contacted sections in the monolithic cavity of LD. The cavity sections which are positively biased provide the optical gain while a negatively biased section behaves as a saturable electroabsorber. Under moderate negative bias applied to the absorber section, the laser exhibits passive mode-locking or Q-switching operation. However with further increasing negative bias, yielding shorter recovery time and larger absorption coefficient, a different emission regime occurs. At threshold of Q-switched lasing operation and under pulsed current pumping, the laser reveals features of SR-like emission. These features have been experimentally observed in multi-section GaAs, AlGaInAs and GaN QW lasers [36-39]. Unfortunately all experimental studies reported in the literature do not distinguish the first emission burst from the subsequent self-pulsations, which can be modulated with a Q-switching pulse envelope. Nevertheless a clear Rabi splitting was observed in the optical spectrum of multi-section LDs when these were expected to produce SR emission [36]. Although there were no reports on experimental or theoretical studies on RNGH instability in a multi-section laser diode, we believe that this subject will receive further attention in future.

## F. The role of carrier population and carrier coherence gratings

In Fig. 14 we study the effect of coherence grating on the gain spectrum (increment) for multimode instability by examining the highest gain frequency $\Omega_{max}$ in Eq.(16) as a function of the pump rate *p*. These numerical studies complement and confirm the outcomes of our analytical studies from Ref.[25].

As a reference, we show the frequency $\Omega_{max}$ obtained by numerically solving the eigen problem (16) with QCL parameters from Table 1 and coherence grating relaxation time $T_{2_g} \neq 0$ (solid blue curve) as well as the calculated Rabi frequency (18) (dashed blue curve) under the same operation conditions. These two reference curves are thus obtained in the presence of mode coupling via scattering on the coherence grating in addition to the coupling via carrier population grating.

The effect of coherence grating can be excluded from Eq.(16) by considering the limit $T_{2_g} \to 0$. We achieve this via reducing $T_{2_g}$ by a factor of $10^{11}$ while maintaining the average decoherence rate ( $T_{2_eff} \neq 0$ ) and the mode coupling via SHB-induced grating of the carrier distribution ($T_g \neq 0$ ).This represents very well the limit of $T_{2_g} \to 0$, when there is no mode coupling via scattering on the coherence grating.

We find that the system (16) still reveals instability and this instability is caused by the SHB effect only. For the pump rates up to p~10, the spectral shape of the instability increment Re($\Lambda$) is similar to the one depicted in Fig.2. (Compare this shape with the warped instability gain curve at much higher pump rate of p~50-60 in Fig. 1.) Thus the instability gain spectra do not indicate a change in the mechanism of multimode instability. In Fig. 14 we add the superscript "(SHB)" in order to distinguish this case and plot the frequency $\Omega_{\max}^{(SHB)}$ calculated numerically from Eq.(16) and $\Omega_{Rabi}^{(SHB)}$ from Eq.(18) for the case when $T_{2_g}$ is reduced by a factor of $10^{11}$ (solid and dashed red curves, respectively).

In Ref. [16], the expression (19) was obtained for the highest increment frequency for the multimode instability caused by the SHB effect, that is in the case we have just discussed above. In Fig. 14 we denote this frequency as $\Omega_{SHB}^{[16]}$ (green curve) and plot it as a function of the pump rate using the parameters of QCL from Table 1.

The following conclusions can be made from comparison between different curves in Fig. 14:

(i) In the presence of coherence grating ($T_{2_g} \neq 0$), the highest instability gain is at the offset frequency $\Omega_{\max}$ which is very close to the Rabi frequency $\Omega_{Rabi}$. Like the last one, $\Omega_{\max}^2$ exhibits a linear growth with the pump rate *p* (solid blue curve in Fig.14). This behavior is a signature of a multimode RNGH-like instability (see the Introduction).

(ii) Without coherence grating ($T_{2_g} \to 0$), both the frequency $\Omega_{\max}$ and the Rabi frequency reduce to $\Omega_{\max}^{(SHB)}$ and $\Omega_{Rabi}^{(SHB)}$, respectively (solid and dashed red curves). However $\left[\Omega_{\max}^{(SHB)}\right]^2$ does not follow anymore the linear growth of $\left[\Omega_{Rabi}^{(SHB)}\right]^2$ and slightly deviates from the Rabi oscillation frequency. Thus even at the pump rate as high as *p*=5 times above the lasing threshold, $\Omega_{\max}^{(SHB)}$ is only by a factor of 1.1 lower than $\Omega_{Rabi}^{(SHB)}$. Therefore we cannot attribute it to a markedly different behavior than the one observed in the presence of additional mode scattering on the coherence grating. Inclusion of the coherence grating just slightly increases the highest instability frequency in the 4×4 submatrix in Eq.(16).

(iii) The frequency $\Omega_{SHB}^{[16]}$ [Eq.(19)] obtained in Ref [16] for the case without coherence grating, that is for case we discussed in (ii), is significantly lower than our frequency $\Omega_{\max}^{(SHB)}$ and the Rabi flopping frequency $\Omega_{Rabi}^{(SHB)}$. As a consequence of such low frequency it was attributed to a multimode instability of a different kind and labeled as a multimode instability caused by the SHB effect. In fact the original system of the coupled mode equations in Ref [16] contains a set of errors. Its adiabatic approximation for slowly varying medium, polarization does not recover the well-known Class-B laser model (See a discussion to our Eqs. (7)-(11) and the Appendix A). As a consequence of these errors, the frequency $\Omega_{SHB}^{[16]} << \Omega_{Rabi}^{(SHB)}$, which is not the case in our model.

In ref [25] we analyze this problem analytically and arrive to the same conclusions. In particular we obtain a closed form expression for the second threshold that indicates that coherence grating just slightly reduces the second threshold. The main source for low-threshold RNGH instability in QCLs is the SHB effect and a set of the relaxation and diffusion time constants that result in $T_g$ being comparable to $T_1$.

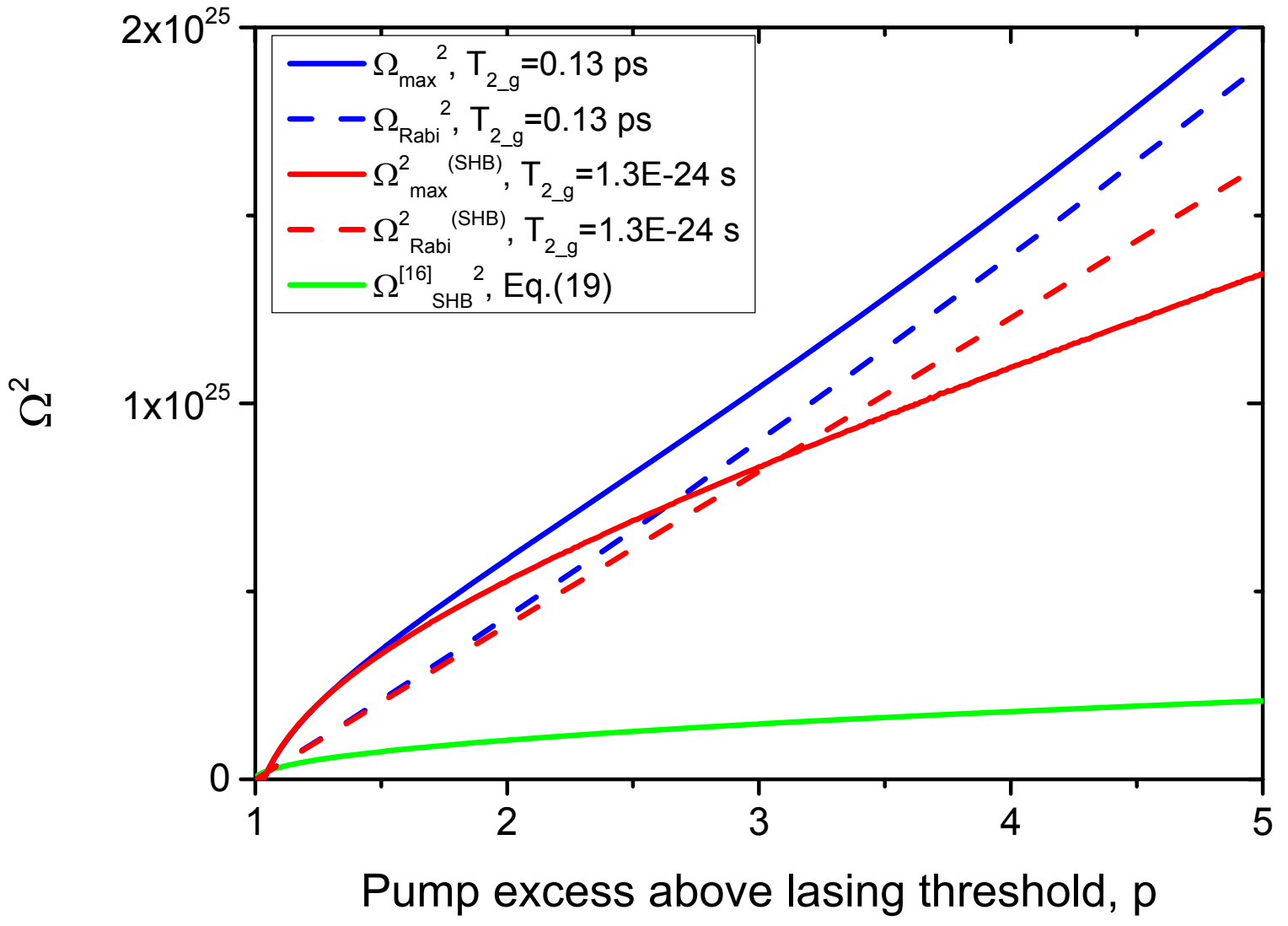

FIG. 14. We plot the squares of the following frequencies as a function of the pump rate, using parameters of QCL from Table 1 ($T_g$=0.927 ps, $T_{2_eff}$=0.14 ps and $T_{2_g}$=0.13 ps unless stated otherwise): the highest instability gain frequency $\Omega_{max}$ calculated numerically from the eigen problem of Eq.(16) ($T_{2_g} \neq 0$, solid blue curve) and corresponding Rabi oscillations frequency (18) (dashed blue curve); the frequency $\Omega_{max}^{(SHB)}$ (solid red curve) and the Rabi frequency $\Omega_{Rabi}^{(SHB)}$ (dashed red

curve) obtained without coherence grating ($T_{2_g}$ is reduced by a factor of $10^{11}$); and finally the frequency $\Omega_{SHB}^{[16]}$ calculated from Eq.(19) (green curve)..

The coherence grating has a much stronger impact on the eigen solutions of the upper 5×5 block in Eq.(16). In Fig. 15 we plot the spectra of the Lyapunov exponents with the largest real parts in the 4×4 (red curves) and 5×5 (blue curves) blocks of Eq.(16) for the two pump rates $p$=9 [Fig.15(a)] and $p$=10 [Fig.15(b)]. Once again we compare cases when the coherence grating is present ($T_{2_g} \neq 0$, solid curves) and when it is absent ($T_{2_g} \to 0$, dashed curves). We observe that without coherence grating (when $T_{2_g} \to 0$) the increment of instability in the 4×4 matrix block (red curves) increases and its spectrum shifts to lower frequencies. It shows similar behavior at the two considered pump rates ($p$=9 and $p$=10). The behavior of the instability increment in the 5×5 block has much in common with the one in 4×4 block but the increment values are shifted down on the vertical axis in Fig. 15 toward negative Lyaponov exponents. Only without coherence grating ($T_{2_g} \to 0$,while $T_{2_eff}$ is unchanged), the increment becomes positive and RNGH instability occurs at $p$=10 (see the behavior of the dashed blue curve in the inset of Fig.15(b)). Note that the RNGH instability does not occur at $p$=9 because $T_{2,eff}/T_1 \neq 0$ [7]. Once the coherence grating is present ($T_{2_g} \neq 0$), the RNGH instability threshold in 5×5 matrix block increases by several times, to around $p$=60 in particular case (See Fig. 1).

In this way it remains only one pleasurable explanation for the low second threshold observed in experiments with QCLs. Namely we conclude that it is related with the instability arising from the 4×4 matrix block in Eq.(16) due to a combined effect of the carrier population and carrier coherence gratings.

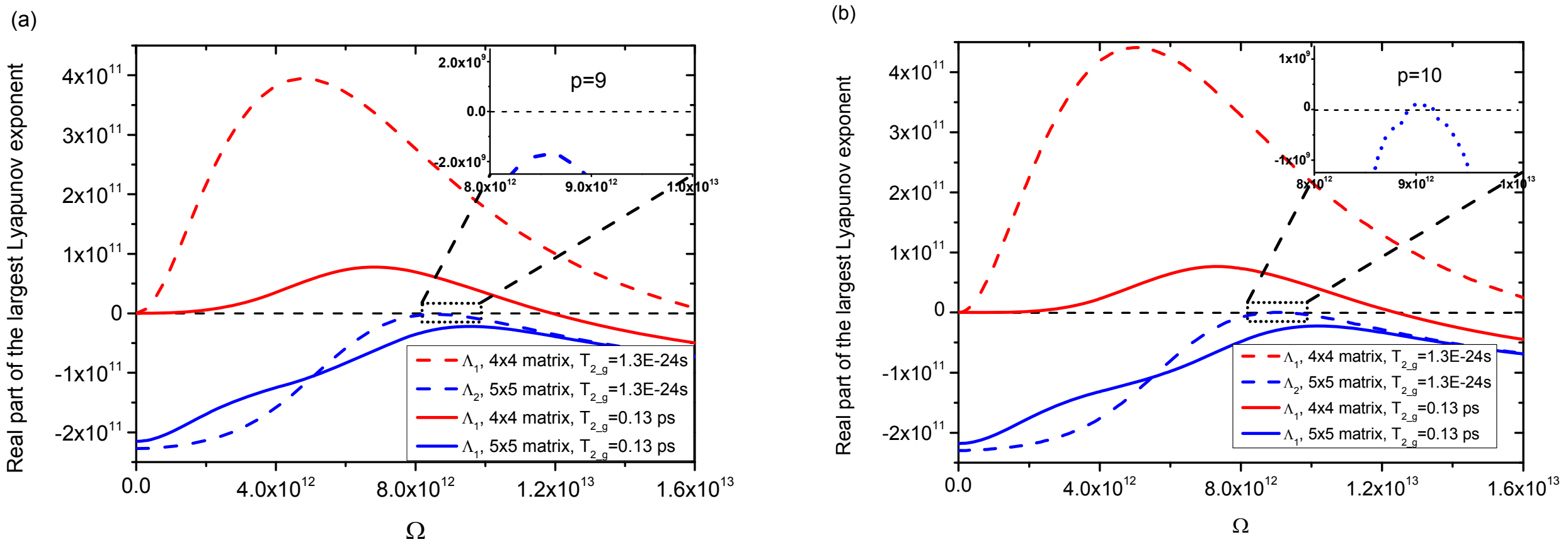

FIG. 15. Spectra of instability increments (largest real parts of the Lyapunov exponents) of the two matrix blocks of 4×4 (red curves) and 5×5 (blue curves) sizes for QCL with parameters from Table 1. The pump rate excess above threshold is $p$=9 (a) and $p$=10 (b) The solid curves stand for the case considered in this paper when we include the coherence grating effects ($T_{2,g}$=0.13ps), while the dashed curves represent the case when the coherence grating relaxation time $T_{2_g}$ is reduced by a factor of $10^{11}$, suppressing all effects caused by the coherence grating. The insets show zooms to the highest RNGH instability increments in the 5×5 matrix block when $T_{2_g} \to 0$ (no coherence grating effects).

## IV. CONCLUSION

In conclusion we have clarified conditions for multimode RNGH instability in a Fabry-Perot cavity laser and proposed a new explanation for low second threshold in QCLs, which is based on a combined effect of the carrier coherence and carrier population gratings. We have studied the impact of the cavity length on the waveform of RNGH self-pulsations and shown that short-cavity QCLs can produce a regular train of ultrashort pulses. There is even a possibility to obtain supperradiant-like emission. Our findings open a practical way of achieving ultra-short pulse production regimes in the mid-infrared spectral range.

Applying the same analysis to conventional laser diodes we find that carrier diffusion in the active bulk semiconductor material or quantum wells prevents the LDs from RNGH self-pulsations.

## ACKNOWLEDGEMENTS

The authors would like to acknowledge the support from the Swiss National Science Foundation (SNF), Ministry of Education, Science and Technological Development (Republic of Serbia), ev. no. III 45010, COST ACTIONs BM1205 and MP1204 as well as the European Union's Horizon 2020 research and innovation programme under the grant agreement No 686731 (SUPERTWIN). DBo is grateful to Vladimir Kocharovsky and Ekaterina Kocharovskaya for valuable discussions.

## APPENDIX

### A. Adiabatic approximation test

In this appendix we consider adiabatic-following approximation that applies to *bidirectional ring lasers* operating on two counter-propagating waves or single-mode *FP cavity lasers.* In these lasers, the standing wave pattern of the field in the cavity leads to the spatial hole burning in the distribution of population inversion. At the same time, our considerations presented below do not apply for a single-frequency *unidirectional ring laser* from Refs. [7,8] because there is no standing wave pattern in the cavity of such laser and associated with it spatial hole burning effect.

The adiabatic-following approximation for the medium polarization is valid when the polarization dynamics is slow and follows instantaneously the optical field in the cavity [9]. For solid state lasers, including semiconductor LDs and QCLs, the dephasing time $T_2$ is much smaller than the relaxation $T_1$. Therefore under adiabatic-following approximation, the behavior of these systems fall into Class-B laser dynamics. These dynamical systems are very well studied, including the cases with mode coupling due to back-scattering and induced population grating. One example can be found in a review [24].

In order to obtain the adiabatic-following approximation for our coupled-mode system (7)-(10), we set the time derivatives in Eqs (8) and (9) to zero, yielding the following instantaneous amplitudes of polarization grating harmonics:

$$\eta_{\pm} = \frac{i\mu}{2\hbar}\left(E_{\pm}\Delta_0 + E_{\mp}\Delta_2^{\mp}\right)T_2 \qquad \text{(A1)}$$

$$\eta_{\pm\pm} = \frac{i\mu}{2\hbar}\left(E_{\pm}\Delta_{\mp}\right)T_{2_g} \qquad \text{(A2)}$$

where for the purpose of this test we have excluded the diffusion terms, assuming that $T_{2_eff} = T_{2_g} = T_2$, and $T_g = T_1$. Substituting these in Eqs. (7), (9) and (10), we find that

$$\frac{n}{c}\frac{dE_{\pm}}{dt} = \mp\frac{dE_{\pm}}{dz} + \frac{NT_{2_eff}\mu^2\Gamma\omega}{2c\hbar n\varepsilon_0}\left(E_{\pm}\Delta_0 + E_{\mp}\Delta_2^{\mp}\right) - \frac{1}{2}l_0 E_{\pm} \qquad \text{(A3)}$$

$$\frac{d(\Delta_0 + \Delta_2^{+}e^{2ikz} + \Delta_2^{-}e^{-2ikz})}{dt} = \\ = \frac{\Delta_{pump}}{T_1} - \frac{\Delta_0 + \Delta_2^{+}e^{2ikz} + \Delta_2^{-}e^{-2ikz}}{T_1} \times \left[1 + a\left((E_-^*E_- + E_+^*E_+) + 2\,\mathrm{Re}(E_-^*E_+e^{-2ikz})\right)\right] \qquad \text{(A4)}$$

where we have introduced the saturation parameter $a = T_1T_2\mu^2/\hbar^2$. This system of equations is in perfect agreement with the well-known Class-B laser model (e.g. see Eqs. (1.10) and (1.11) in Ref. [24]). In contrast to our coupled-mode expansion, the adiabatic approximation that follows from the equations (8)-(10) of Ref. [16] disagrees with the well-established coupled-mode model for a Class-B laser.